\documentclass[useAMS,usenatbib]{mnras}
\def\aap{A\&A}
\def\apj{ApJ}
\def\apjs{ApJS}
\def\apjl{ApJL}
\def\mnras{MNRAS}

\def\aaps{A\&A Supp.}
\def\prd{Phys.Rev.D}         % Physical Review D
\def\pasj{PASJ}              % Publications of the ASJ
\def\apss{ApSS}
\def\jcap{JCAP}

\usepackage{amsmath}
\usepackage{natbib}
\usepackage{epsfig}
\usepackage{txfonts}
\usepackage{graphicx}% Include figure files
\usepackage{dcolumn}% Align table columns on decimal point
\usepackage{amsfonts}
\usepackage{amssymb}
\usepackage{bm}% bold math
\usepackage{url}
\usepackage{subfigure}
\usepackage{txfonts}
\usepackage{color}
\usepackage{enumitem}
\usepackage{soul}
\usepackage{float}

\title[Pressure fluctuations in Coma cluster]{
  Thermal SZ fluctuations in the ICM: probing turbulence and
  thermodynamics in Coma cluster with {\it Planck}}
\author[Khatri \& Gaspari]{Rishi Khatri$^{1}$\thanks{E-mail: khatri@theory.tifr.res.in} 
and Massimo Gaspari$^{2}$\thanks{E-mail: mgaspari@astro.princeton.edu; {\it Einstein} and {\it Spitzer} Fellow}\\
$^{1}$Department of Theoretical Physics, Tata Institute of Fundamental Research, Homi Bhabha Road, Colaba,
Mumbai 400005 India\\
$^{2}$Department of Astrophysical Sciences, Princeton University, 4 Ivy Lane, Princeton, NJ 08544-1001
USA
}
\begin{document}

\newcommand{\id}{{{\rm d}}}
\newcommand{\expe}{{{\rm e}}}

\newcommand{\Acmb}{{{A_{\rm CMB}}}}
\newcommand{\Alf}{{{A_{\rm lf}}}}
\newcommand{\dellf}{{{\delta_{\rm lf}}}}
\newcommand{\delblf}{{{\delta^{\rm lf}_{\beta}}}}
\newcommand{\flf}{{{f_{\nu}^{\rm lf}}}}
\newcommand{\nulf}{{{\nu_{0}^{\rm lf}}}}
\newcommand{\blf}{{{\beta_{\rm lf}}}}
\newcommand{\Adst}{{{A_{\rm d}}}}
\newcommand{\deldst}{{{\delta_{\rm d}}}}
\newcommand{\delbdst}{{{\delta^{\rm d}_{\beta}}}}
\newcommand{\fdst}{{{f_{\nu}^{\rm d}}}}
\newcommand{\nudst}{{{\nu_{0}^{\rm d}}}}
\newcommand{\bdst}{{{\beta_{\rm d}}}}
\newcommand{\Tdst}{{{T_{\rm d}}}}
\newcommand{\Aco}{{{A_{\rm co}}}}
\newcommand{\fco}{{{f_{\nu}^{\rm co}}}}
\newcommand{\kB}{{{k_{\rm B}}}}
\newcommand{\hbn}{\mathbf{\hat{n}}}
\newcommand{\sigT}{{{\sigma_{\rm T}}}}
\newcommand{\me}{{{m_{\rm e}}}}
\newcommand{\Ne}{{{n_{\rm e}}}}
\newcommand{\Te}{{{T_{\rm e}}}}
\newcommand{\bx}{{{\mathbf{x}}}}
\newcommand{\br}{{{\mathbf{r}}}}
\newcommand{\bqp}{{{\mathbf{q'}}}}
\newcommand{\bq}{{{\mathbf{q}}}}
\newcommand{\bk}{{{\mathbf{k}}}}
\newcommand{\bkp}{{{\mathbf{k'}}}}
\newcommand{\mcP}{\mathcal{P}}
\newcommand{\bkz}{{{\mathbf{k_z}}}}
\newcommand{\bkt}{{{\mathbf{k_{\theta}}}}}
\newcommand{\bktp}{{{\mathbf{k_{\theta}'}}}}
\newcommand{\kt}{{{{k_{\theta}}}}}
\newcommand{\ktp}{{{{k_{\theta}'}}}}
\newcommand{\bhz}{{{\mathbf{\hat{z}}}}}
\newcommand{\bt}{\boldsymbol{\theta}}
\newcommand{\changeR}[1]{\textcolor{red}{#1}}

%MAX defs
\def\lta{\stacksymbols{<}{\sim}{2.5}{.2}}
\def\gta{\stacksymbols{>}{\sim}{2.5}{.2}}
\def\approxprop{\stacksymbols{\propto}{\sim}{3}{.5}}
\newcommand\msun{\rm{M_{\odot}}}

\maketitle

\begin{abstract}
We report the detection of thermal Sunyaev-Zeldovich (SZ) effect  fluctuations in the intracluster medium (ICM) of Coma cluster observed with {\it Planck}. 
The SZ data links the maximum observable X-ray scale to the large Mpc scale, extending our 
knowledge of the power spectrum of ICM fluctuations.
Deprojecting the 2-d SZ perturbations into 3-d pressure fluctuations, we
find an amplitude spectrum which peaks at $\delta P/P =
  33\pm 12\%$ and $74\pm19\%$ in the $15'$ and $40'$ radius region, respectively.
We perform tests to ensure fluctuations are intrinsic to the cluster and not due to noise contamination.
By using high-resolution hydrodynamical models, we improve the ICM turbulence constraints
in Coma, finding 3-d Mach number ${\rm Ma_{3d}}= 0.8\pm0.3$ (15' region), increasing to supersonic values at larger radii (40'), and an injection scale $L_{\rm inj}\approx 500$ kpc. 
Such properties are consistent with driving due to mergers, in particular tied to internal galaxy groups.
The large pressure fluctuations show that Coma is in adiabatic mode (mediated by sound waves), rather than isobaric mode (mediated by buoyancy waves). 
As predicted by turbulence models, the distribution of SZ fluctuations is log-normal with mild non-Gaussianities (heavy tails).
The substantial non-thermal pressure support implies hydrostatic mass bias 
$b_M=-15\%$ to $-45\%$ from the core to the outskirt region, respectively.
While total SZ power probes the thermal energy content, the SZ fluctuations constrain the non-thermal deviations important for precision cosmology.
The proposed, novel approach can be exploited by multifrequency
observations using ground based interferometers and future space CMB missions. 
\end{abstract}

\section{Introduction}
The {\it Planck} cosmic microwave background (CMB) mission
  \citep{planckmission} has for the first time allowed us to retrieve
  all-sky maps \citep{planckymap,hs2014,lily} of the Sunyaev-Zeldovich effect
  (SZ; \citealt{zs1969}). The chief advantage of Planck over other
past and current  satellite and ground based missions is its multiple
frequency channels covering the low frequency Rayleigh-Jeans  as well as the higher
frequency Wien region of the CMB
spectrum making the separation of the thermal Sunyaev-Zeldovich from CMB
and foregrounds feasible. As the blackbody photons from the CMB interact with hot
  intergalactic/intracluster medium (IGM, ICM) traveling from the last
  scattering surface  (\citealt{sz1970c}; \citealt{Peebles1970}) to us, Compton scattering upscatters a fraction of photons to higher
  energy creating a distortion from the  Planck spectrum $I_{\nu}^{\rm
    Planck}$ \citep{sz1972}. The change in the
  intensity $\Delta I_{\nu}=I_{\nu}-I_{\nu}^{\rm Planck}$ of the CMB
  radiation is given by \citep{zs1969}
\begin{align}
\Delta I_{\nu}=y\frac{2 h\nu^3}{c^2}\frac{x
  \expe^x}{(\expe^x-1)^2}\left[x\left(\frac{\expe^x+1}{\expe^x-1}\right)-4\right],
\label{Eq:y}
\end{align}
where $x=\frac{h\nu}{\kB T}$, $T=2.725(1+z)$ is the CMB temperature at
redshift $z$, $\nu=\nu_0(1+z)$ is the frequency of CMB photon at
  redshift $z$,  $\nu_0$ is the observed frequency today ($z=0$),  $h$ is the   Planck's constant, $\kB$ is
Boltzmann constant and $c$ is the speed of light. The amplitude of the
distortion, $y$, is proportional to the integral of the pressure $P$ along the line
of sight,
\begin{align}
y=\frac{\sigT}{\me c^2} \int\id s\, \Ne  \kB \Te,
\end{align}
where $\Te$ and $\Ne$ are the electron temperature and
  electron number density respectively in the ICM plasma, $\me$ is the mass
of the electron, $\sigT$ is the Thomson scattering cross section, and $s$ is the distance coordinate along the
line of sight.

The Sunyaev-Zeldovich amplitude, $y$, contains information
  about the properties of the galaxy cluster which in turn are sensitive to the
  cosmological parameters. The {\it Planck} collaboration as well as independent
  groups have used the $y$ maps created using the {\it Planck} data to
  study the clusters themselves as well as to measure the cosmological
  parameters (e.g., \citealt{hs2014,planckymap,planckclusters2015,cosmocluster2015,rma2015})
  %planckclphy1,planckclphy2,planckclphy4,ebt2015}. 
  In particular, the {\it Planck} collaboration has used the
  $y$ parameter measurements of the Coma and Virgo cluster to constrain the average
  properties of these clusters \citep{planckcoma,virgo2015}. Since these
  nearby clusters are well resolved by {\it Planck}, detecting $y$
  signal out to several Mpc, it opens up the possibility of studying the
  $y$ fluctuations -- and hence the ICM pressure perturbations --
  on large scales, complementing the X-ray studies focusing on smaller scales 
  (e.g., \citealt{Schuecker:2004}; \citealt{churazov2012}; \citealt{Sanders:2012}; \citealt{Gaspari:2013,Gaspari:2014,Hofmann:2016} and references within). The smallest scale we
  can study with {\it Planck} is limited by the angular resolution of $10'$
  of the 100 GHz channel. {We need the 100 GHz channel to be able
    to do component separation and minimize contamination from the other components. }

While X-ray telescopes such as {\it Chandra} and {\it XMM-Newton} 
provide us with extensive details on the density and temperature of the ICM
plasma ($T_{\rm x}\approx1$-$10$\, keV), the X-ray emissivity is $\propto
n_{\rm e}\,n_{\rm ion}$, where $n_{\rm ion}$ is the ion number density, i.e., limited to the core region due to the steep negative density gradient. The SZ signal can instead be detected with significance at much larger radii, granting the complementary view to X-ray observations. On the other hand, it should be noted that the temperature decline limits the SZ signal ($\propto n_{\rm e}\,T_{\rm e}$) at large radii too (\citealt{Hallman:2007}).
Using high-resolution 3-d simulations, \citet{Gaspari:2013} and
\citet{Gaspari:2014} have shown that the plasma perturbations in all the
thermodynamic variables (density, entropy, pressure; $\rho, K, P$) are
tightly related to the dynamical level of the cluster, i.e., how strong the turbulent motions are.
In particular, the peak of the Fourier amplitude spectrum, $A_{\rho}(k_{\rm peak})$, 
of the relative {\it density} perturbations, $\delta \rho/\bar{\rho}$ is linearly tied to the 3-d Mach number of gas motions as (\citealt{Gaspari:2013})
\begin{align}
{\rm Ma_{\rm 3d}}\approx4\,A_{\rho}(k_{\rm peak})\approx 2.4\,A_{P}(k_{\rm peak}),\label{Ma_rho}
\end{align} 
where ${\rm Ma_{3d}}=\sqrt{3}\,{\rm Ma_{1d}}=\sigma_v/c_{\rm s}$,
$\sigma_v$ is the turbulent velocity dispersion, $c_s$ is the ICM adiabatic
sound speed and ${\rm Ma_{1d}}$ is the 1-d Mach number. In the last
equality we have assumed adiabatic perturbations with adiabatic index
$\gamma=5/3$ and $A_P$ is the amplitude of pressure fluctuations  $\delta P/\bar{P}$.
The cascade of the spectrum is instead related to the microphysics of the plasma. 
If  thermal conduction (diffusion of internal energy via plasma electrons) is dominant, 
the density (or temperature) amplitude spectrum becomes steeper than that of the velocity field by up to factor of 5 (Fig.~2 in \citealt{Gaspari:2014}). For weaker diffusion, both slopes tend to the Kolmogorov power spectrum, $P_k\propto k^{-11/3}$ ($A_k\propto k^{-1/3}$). 
However, the density and velocity slopes do not follow each other linearly at progressively smaller scales, as thermodynamic quantities are not pure passive scalars. The simple analytic conversion in Eq.~\ref{Ma_rho} between velocity and thermodynamic fluctuations should be thus used only near the injection scale (spectrum peak), i.e., between total variances.

In this paper, we will for the first time use the Compton $y$ fluctuations, $\delta y/y$, to retrieve pressure fluctuations $\delta P/P$, and thus assess the level of gas motions at scales larger than that probed via X-ray surface brightness in Coma galaxy cluster (\citealt{churazov2012,Gaspari:2013}).
We will show that substantial {\it pressure} fluctuations ($\delta
P/\bar{P}\simeq\gamma\,\delta\rho/\bar{\rho}$) of the order $30$ percent 
are present at several 100 kpc, while gradually fading at radii $r\gtrsim 1$\,Mpc.
This implies gas turbulent motions\footnote{Turbulence should be here
  interpreted as large scale eddies, sometimes referred to as `bulk
  motions' given the long turnover timescale.} with Mach number ${\rm
  Ma_{3d}}\approx0.8$, i.e., subsonic yet significant turbulence, as the average sound speed of Coma is $c_{\rm s}\approx1500$\,km\,s$^{-1}$.
This is consistent with motions driven by mergers and cosmological flows
(e.g., \citealt{Lau:2009, Vazza:2009, Vazza:2011}; \citealt{Miniati:2014,Battaglia:2015,Schmidt:2016}).

It is crucial to understand the physics of the ICM, as hot gaseous haloes have been found to be ubiquitous from massive galaxy clusters to elliptical and spiral galaxies (e.g., \citealt{Anderson:2015}).
Remarkably, the cascading large-scale turbulence in the hot halo can also
drastically alter the accretion mode onto the central supermassive black
hole, igniting `chaotic cold accretion' (e.g.,
\citealt{Gaspari:2015_cca}). 
The presence of significant non-thermal pressure support has important
implications not only for the diffuse gas physics, but also for
cosmology.   Being the largest collapsed structures, galaxy clusters
are highly sensitive to the cosmological parameters.
To achieve high-precision cosmology, however, we require precise estimates
of the cluster masses, minimizing errors and scatter. The usual approach of
using the average SZ and X-ray signal, which can only give an estimate of
thermal pressure support, gives a biased estimate of the cluster mass. This
bias, which comes from the non-thermal pressure contributions {(assuming
hydrostatic equilibrium\footnote{Part of the bias may come
  from deviations from spherical symmetry or failure of hydrostatic
  equilibrium assumption.}
  )}, has been typically estimated by using numerical simulations
%\citep{dolag2005,rasia2006,piff2008,iap2008,maier2009,Vazza:2009,Shaw:2010,mene2010,Vazza:2011,iap2011,conte2011,bbps2012,krause2012,nelson2014,Biffi:2016}.
(e.g., \citealt{dolag2005,Vazza:2011,iap2011,bbps2012,nelson2014,Biffi:2016} and references within).
The non-thermal pressure has also been studied using analytical methods
guided by both simulations and observations (e.g., \citealt{cnm2012,cmn2013,shi2015}). 
Observationally the evidence for the non-thermal pressure support
is indirect via the discrepancy between the X-ray, SZ or dynamical masses and
the masses derived via weak or strong lensing (e.g., \citealt{mahdavi2008,zhang2010,linden2014,smith2016}). 
{Radio observations can constrain the magnetic field strength via Faraday rotation measure, which is typically a few $\mu$G (e.g., \citealt{Ferrari:2008,Bonafede:2010}), and thus can estimate the magnetic pressure role.
The SZ fluctuations open up the possibility to directly probe the turbulence contribution
to the non-thermal pressure support, which we expect to be dominant from cosmological simulations (e.g., \citealt{Avestruz:2016}).}

\begin{figure*}[!ht]
\centering
\includegraphics[scale=0.26]{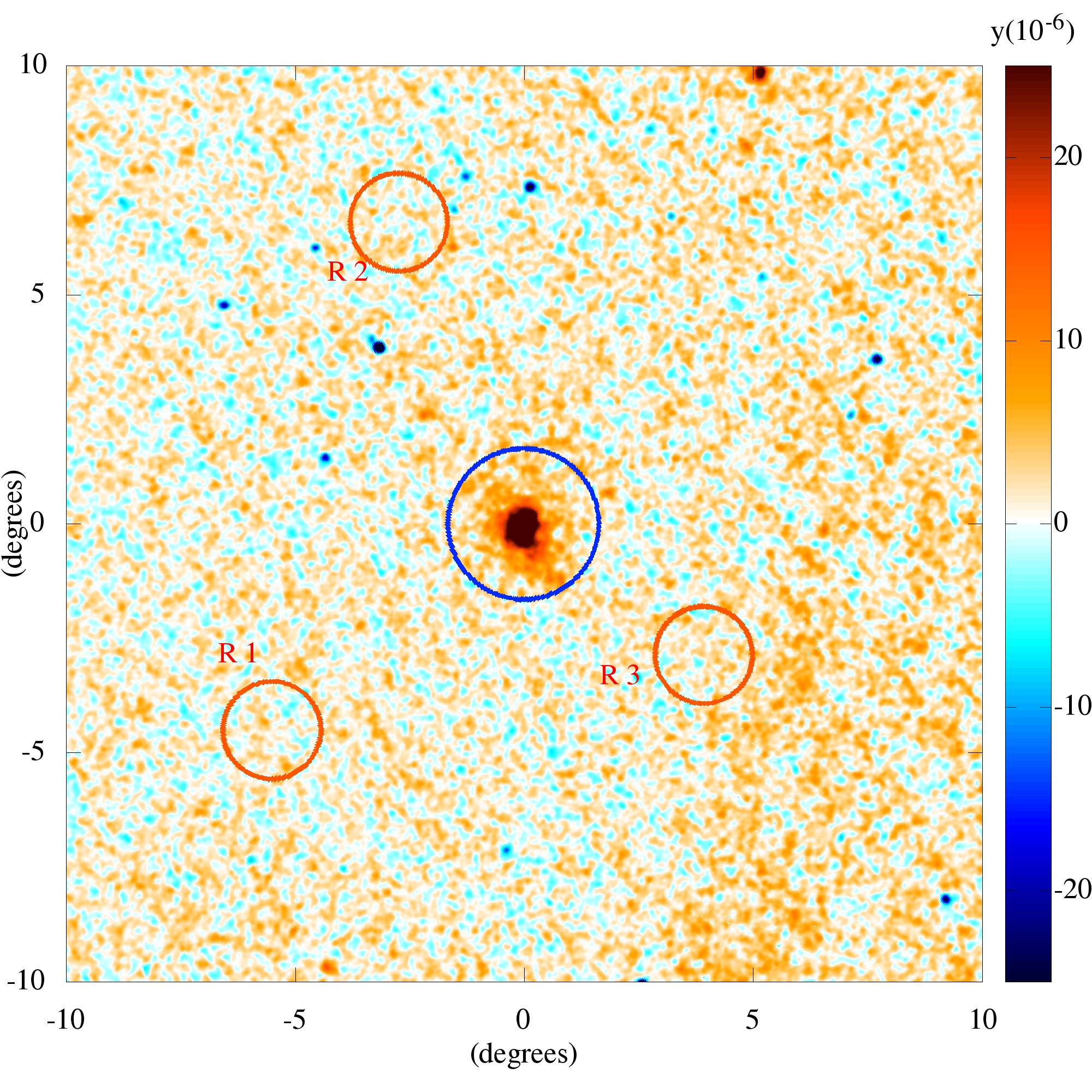}
\caption{\label{Fig:map}We show $20^{\circ}\times 20^{\circ}$ region
  centered on the Coma cluster in LIL map. The $R_{200}\approx 2.62~{\rm Mpc}$ region
  is marked with a circle at the center of the map. The three $1^{\circ}$ regions
  used to test for typical contamination in this part of the sky are also
  marked with circles and labeled $R1,R2,R3$. The bright cluster at the top-right
  edge of the plot is ACO 1795.
}
\end{figure*}

For the first exploratory study, we choose Coma cluster.
Coma (or Abell 1656) is one of the biggest clusters in angular size
    visible to us with highest signal to noise ratio in {\it Planck} data and in a
    relatively uncontaminated part of the sky. It
    is thus the ideal object to start looking for $y$-fluctuations
    which are expected to be weaker than the average $y$ signal. The
    Coma cluster is situated at a redshift of $z=0.0232$ \citep{comacluster}. The
    physical distance depends on the cosmological parameters and we
    take the 
    distance of the Coma cluster to be 93 Mpc (choosing the same cosmology
    as in \citealt{churazov2012}). At this distance $10' = 270.5
    ~{\rm kpc}$ and $40' \approx 1~{\rm Mpc}$. A crude estimate of the virial radius is $R_{200}\approx2\,R_{500}$, where
    $R_{500}=1.31$\,Mpc. Coma is classified as a hot, massive galaxy
    cluster ($M_{\rm vir}\approx10^{15}\,\msun$) with average core plasma
    temperature 8.5\,keV and electron density $n_{\rm e}\simeq4\times10^{-3}
      ~{\rm cm^{-3}}$, which
    decreases to $10^{-4}$\,cm$^{-3}$ at  $R_{500}$ (cf.~{\it XMM} profiles in
    \citealt{Gaspari:2013}, Fig.~1).  Coma is
    an archetypal non-cool-core cluster with radiative cooling time greater than
    the Hubble time.    
The main objective of the current investigation is finding the level of SZ
fluctuations and SZ power spectrum of Coma cluster, carefully
assessing the role of projection and contamination, and to provide the first exploratory analysis which can be leveraged 
and extended by future SZ surveys.
The paper is organized as follows.
In \S2, we describe how to retrieve SZ fluctuations from {\it Planck} data on top of the average $y$ profile.
In \S3, we deproject the $y$ (2-d) power spectrum into the 3-d power spectrum of
pressure perturbations, $\delta P/P$.
In \S4, we discuss the key physical implications arising from SZ fluctuations, focusing on ICM turbulence constraints, thermodynamics, and the mass bias related to non-thermal pressure.
In \S5, we summarize the main conclusions and discuss future prospects. 

\section{SZ fluctuations in Coma cluster}
The {\it Planck} collaboration has released the Sunyaev-Zeldovich effect maps \citep{planckymap}
calculated from two different algorithms NILC (Needlet Internal Linear
Combination) \citep{nilc} and MILCA (Modified Internal Linear Combination
Algorithm; \citealt{milca}). Several independent groups have also used {\it Planck}
data to construct the SZ effect maps \citep{hs2014,lily}. We
will use the MILCA and NILC maps as well as the SZ map  from
\citet{lily} based on pixel based parameter fitting algorithm LIL
(Linearized Iterative Least-squares) developed in \citet{k2014}. The main
difference between the {\it Planck} collaboration maps and our LIL map is that in
MILCA/NILC, which use internal linear combination methods, the local
monopole is removed while LIL preserves the monopole or the average SZ
signal. In addition, the ILC methods determine the contamination
  spectrum from the data separately in different parts of the sky and try to remove all
  contamination components including the CO emission. The LIL method on the
  other hand ignores the CO emission and fits for a parametric
  contamination model in each pixel in the sky. This is not a
    problem as the CO emission is negligible in the Coma
  region of the sky. We refer to \citet{lily} for
  detailed comparison of LIL with NILC/MILCA algorithms.
We will use all three, NILC, MILCA and LIL, maps for our analysis. As we will see, all three methods agree and
choosing one map over the other does not affect our results and conclusions. 
The three maps, calculated with different algorithms, have
  different levels of contamination from other components. The comparison
  of LIL maps with NILC/MILCA maps serves as a test against non-SZ contamination.
\begin{figure}
\resizebox{\hsize}{!}{\includegraphics{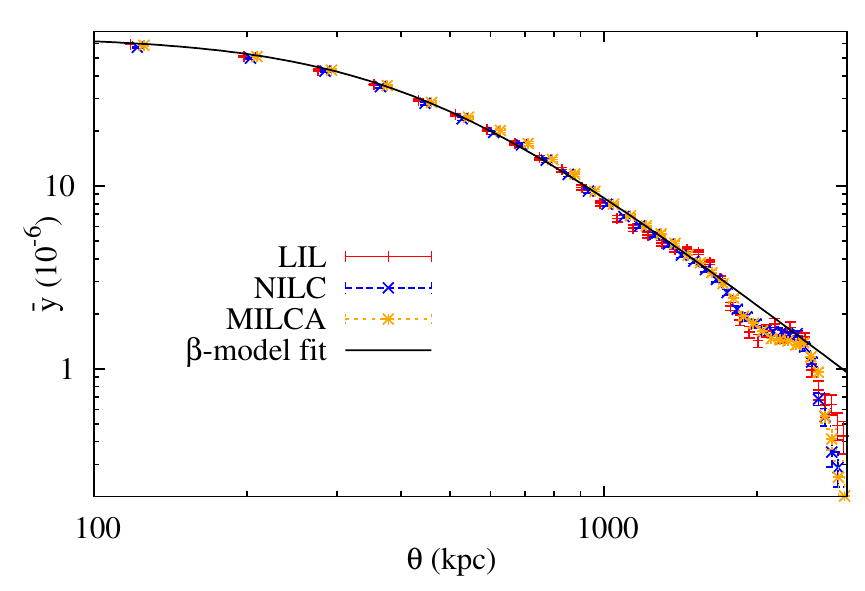}}
\caption{\label{Fig:comaprofile}The measured $y$ profile of Coma cluster in
different $y$ maps (Sec.~2) and the best fit $\beta$-model profile to the LIL
map. The points for LIL and MILCA  have been slightly offset from the
center of the bin in the x
direction to make them easily distinguishable. The distance to Coma of
93 Mpc is used. The size of a bin is $3'$ or $81 ~\rm{kpc}$.
}
\end{figure}

The average SZ and pressure profiles and related properties of the Coma cluster have been
studied in detail in \citet{planckcoma}. We are interested in the
fluctuations in the SZ signal and related ICM pressure. To study the fluctuations
in SZ, $\delta y(\bt)/\bar{y}(\theta)$,
we want to remove the average
profile $\bar{y}(\theta)$, where $\bt$ is the two dimensional
position vector in the plane of the sky 
and $\theta$ is its magnitude or the distance from the center of the cluster. 
We assume spherical symmetry,
so that $\bar{y}$ depends only on the radial distance
$\theta$.
As can be seen in Fig. \ref{Fig:map}, such an assumption is a fair approximation.
 The 2-d to 3-d deconvolution (Abel transform) can be done
exactly only under the assumption of spherical symmetry since the projection
invariably leads to loss of information. Although there have been studies to
correct for the asphericity when doing a statistical study involving large
number of clusters \citep[e.g.][]{fp2002}, such an attempt for a single
cluster would be highly sensitive to the model of asphericity used. This is
especially clear from Fig. \ref{Fig:map} where it can be seen that  the departure
from  sphericity is non-trivial and it is a matter of choice what part
of the anisotropy we
assign to the departure of  average model from sphericity, i.e. the static
background,  and what part we assign to dynamical processes such as mergers
and turbulence.

\begin{figure}
\centering
\resizebox{\hsize}{!}{\includegraphics{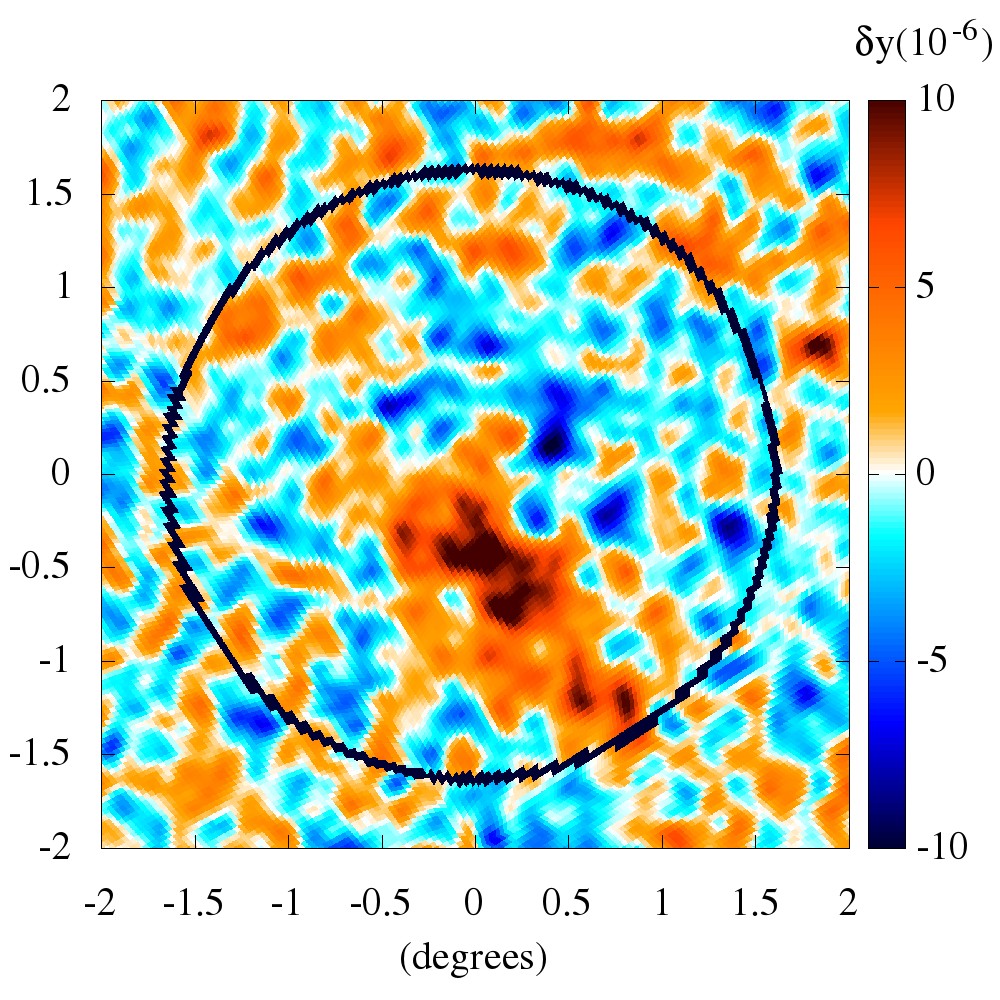}}
\caption{\label{Fig:comayfluct}Fluctuations (LIL map) of y, $\delta y(\bt)$ in Coma cluster with the average
  profile, estimated in rings of 3 arcmin and linearly interpolated,
  subtracted. The circle marks $R_{200}\approx 2.62~{\rm Mpc}$ region
 of the Coma cluster.
  }
\end{figure}

\begin{figure}
\centering
\resizebox{\hsize}{!}{\includegraphics{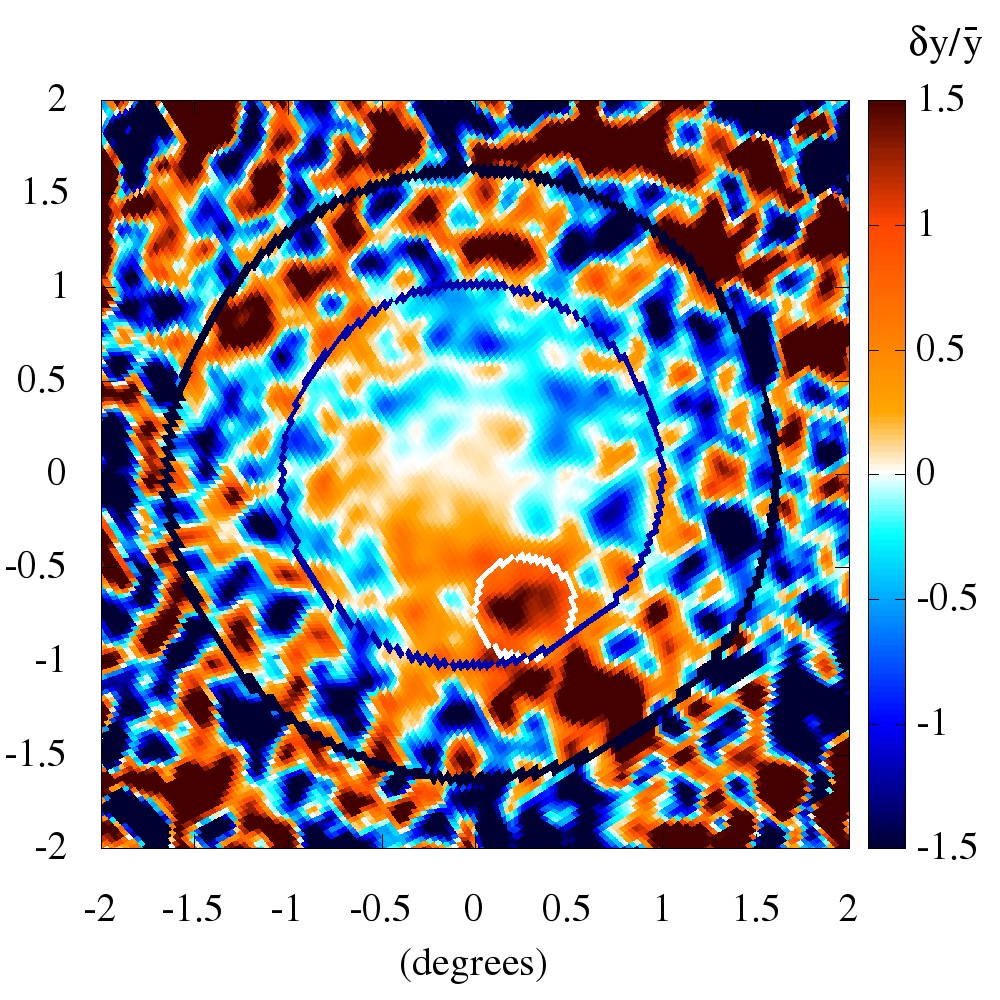}}
\caption{\label{Fig:comayfluctnormed}Fractional fluctuations of $y$ (LIL
  map), $\delta y(\bt)/\bar{y}$ in Coma cluster. The outer circle marks $R_{200}\approx 2.62~{\rm Mpc}$ region
 of the Coma cluster. The inner blue circle marks $60'$ radius. {The white
 circle with radius $15'$ encloses the group of galaxies NGC 4839.} 
 The fluctuations become nonlinear as we move away from the the center of the cluster.
}
\end{figure}

We extract the average profile, $\bar{y}$, by averaging in rings
of width 3 arcmin around the SZ peak in the Coma cluster at galactic
coordinates ($l=57.3^{\circ},b=87.99^{\circ}$)\footnote{This is
  approximately half a pixel ($\sim 0.5'$)
  away from the center used by {\it Planck} collaboration.
  Such a difference is much smaller than the map resolution and does not affect the results.}. The profile is shown in
Fig. \ref{Fig:comaprofile} for NILC, MILCA and LIL maps.
Since LIL also has monopole contribution from background SZ signal, we
have subtracted an estimated $y_{\rm background}=10^{-6}$ from the profile
\citep{ks2015}. 
The $\beta$-model has been found to be a good fit to the SZ profile in South Pole Telescope (SPT)
clusters  \citep{Plagge:2010}. Defining
the $\beta$-model as
\begin{align}
\bar{y}_{\beta}(\theta)=\frac{y_0}{\left(1+\theta^2/\theta_c^2\right)^{\beta}}\label{Eq:yprof}
\end{align}
we get the best fit values to the LIL profile at $\theta<1200~{\rm kpc}$ of $y_0=(65.5\pm 0.31)\times
10^{-6},\,\theta_c=(419.6\pm 9.26)~{\rm kpc},\,\beta=1.07\pm 0.024$ where $\theta$ is the
radial distance from the center of the cluster in the 2-d plane of the sky.
At $\theta \gtrsim 1200 ~{\rm kpc}$ we see a deviation from the
$\beta$-profile. In particular, the flattening in the observed
profile at $\sim 2000~\rm{kpc}$ coincides with the recently reported identification of
a shock in the SZ map of Coma at a distance of $\sim$\,75' from the center
by \citet{ebt2015}. {We note that this feature is absent from the
  profile reported by \citet{planckcoma}. However the radial bins used by
  \citet{planckcoma} are much larger than the size of this feature and it
  is most  likely averaged out in their profile.}
For comparison\footnote{X-ray studies use a different $\beta'$ definition starting from the density profile
$\rho\propto(1+r^2/r_{\rm c}^2)^{3\beta'/2}$. The projected {\it isothermal} profile for $y$ would then be
$\propto(1+\theta^2/\theta^2_{\rm c})^{1/2-3\beta'/2}$, i.e., $\beta'=(2\beta+1)/3\simeq1.05$ for Coma. 
the X-ray brightness profile is instead $\propto(1+\theta^2/\theta^2_{\rm c})^{1/2-3\beta'}$, which is steeper as X-ray emissivity $\propto \rho^2$. Coma X-ray data fits $\beta'\simeq0.75$; such discrepancy with SZ arises because the isothermal assumption is violated outside the core.}, 
the X-ray surface brightness profile of Coma has core radius of $272$\,kpc and steeper slope, ${\rm SB}_{\rm x}\propto (1+\theta^2/\theta^2_{\rm c})^{-1.75}$; \citet{Gaspari:2013}.
If we assume
spherical symmetry, we can get the pressure profile from the SZ profile
using Abel transform, 
\begin{align}
\bar{P}(r)&=\frac{\me c^2}{\sigT}\frac{1}{\pi}\int_{\infty}^{r}\frac{\id y}{\id \theta}\frac{\id
  \theta}{\left(\theta^2-r^2\right)^{1/2}}\nonumber\\
&=\frac{P_0}{\left(1+r^2/r_c^2\right)^{\beta+1/2}},\label{Eq:pprof}
\end{align}
where $r$ is the radial distance from the center in 3-d,
$r^2=\theta^2+z^2$, $z$ is the distance along the line of sight direction and 
\begin{align}
P_0=\frac{\me c^2}{\sigT}\frac{y_0}{r_c}\frac{\Gamma(\beta+0.5)}{\sqrt{\pi}\Gamma(\beta)},
\end{align}
where $\Gamma$ is the usual gamma function.

 \begin{figure}
\resizebox{\hsize}{!}{\includegraphics{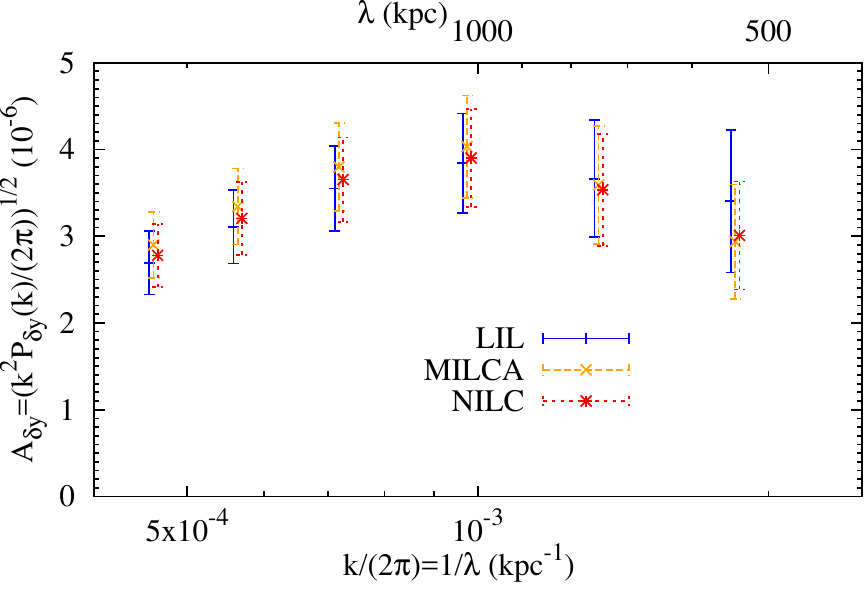}}
\caption{\label{Fig:szfluc}Power spectrum of fluctuations of y, $\delta
  y=y-\bar{y}$, in Coma with the average
  profile subtracted for MILCA,NILC and LIL
  maps. The NILC data point is at the center of the $k$ bin while the other
data points are slightly offset to make them distinguishable from each other.}
\resizebox{\hsize}{!}{\includegraphics{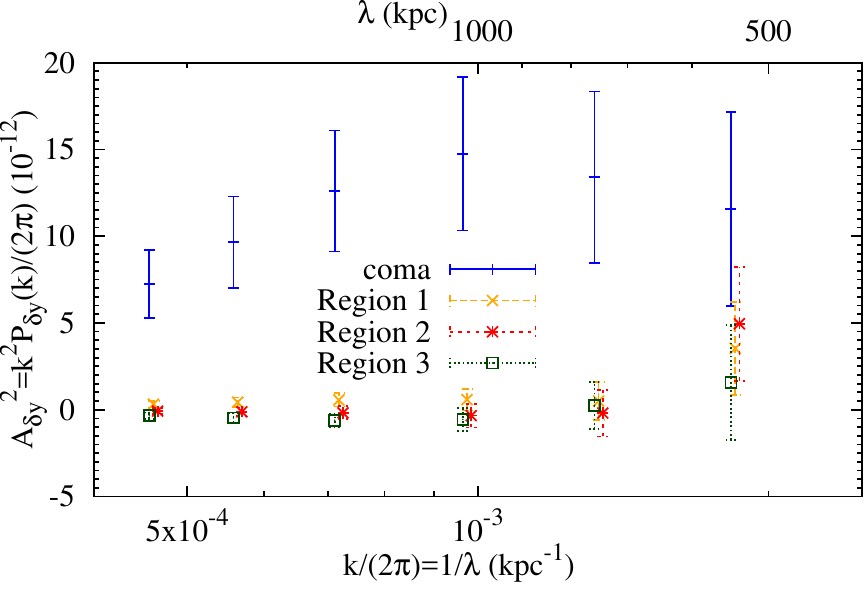}}
\caption{\label{Fig:testfluc}Power spectrum of fluctuations of $y$, $\delta
  y=y-\bar{y}$, in Coma
  and three test regions around Coma in LIL map. Note that the cross power
  spectrum between the two half ring maps in the test regions is in fact negative for all
  points except the last indicating that these test regions are dominated by
  noise. The errors are dominated by sample variance and are
    therefore proportional to the power spectrum on large scales for same
    binning. } 
\end{figure}

We will use the $\beta-$model fit of the profile to just simplify the calculation
of the window functions that will come up when we calculate the
fluctuations. To extract the final relative perturbations, 
we subtract instead the actual {\it measured} average profile $\bar{y}$,
linearly interpolated between the bins
from the total signal $y$  to get the map of $y$ fluctuations, $\delta y(\bt)=y(\bt)-\bar{y}$.
This way we are not dependent on the arbitrary choice of an idealized model (e.g., $\beta$- or Arnaud profile). 
The map of $\delta y$ is shown in Fig. \ref{Fig:comayfluct}
in a $2^{\circ}\times 2^{\circ}$ region around the Coma cluster for LIL
map and the fractional fluctuations $\delta y/\bar{y}$ are shown in Fig. \ref{Fig:comayfluctnormed}.
The $\beta$-model fit will however be useful when we test for contamination in
our results.

We will calculate the angular power spectra using the publicly available
software PolSpice \citep{ps1,ps2} which also calculates the covariance matrix. The
calculation is similar to \citet{hivon2002} and \citet{xspect} but is done in
correlation space instead of harmonic space which makes the
  deconvolution of masks much faster.
We convert the angular power spectrum to Fourier power spectrum using the flat
sky approximation \citep{jk1998}.  See Appendix \ref{appa} for a short derivation and
also for the definition of our Fourier transform convention which differs 
from that in \citet{churazov2012,Gaspari:2013}. The {\it Planck} collaboration
provides half-ring maps which have the same signal as well as almost the same
noise amplitude but the noise between them is uncorrelated. We will use the cross-power
spectra of half-ring y-maps so that the uncorrelated white noise is automatically
canceled. All the power spectra in this paper have been corrected
  for the effect of the masks and the $10'$ FWHM (full width half maximum)
  beam of the SZ maps.

We show the power spectrum of the Coma SZ fluctuations, $\delta y=y-\bar{y}$ in LIL, MILCA and NILC
maps in Fig. \ref{Fig:szfluc} calculated after masking the rest of
  the sky. The amplitude
$A_{\delta y}$ of the power spectrum $P_{\delta y}$ is plotted. The mask is a map consisting of
$1$s in the region occupied by Coma and $0$s everywhere else which leaves $60'$
radius region at the position of the Coma cluster unmasked. 
We apodize the mask before multiplication with the $y$-map to
  avoid high frequency artifacts in power spectrum due to the sharp edge in the
  mask. The apodization is done using a $15'$ Gaussian replacing the $1$s  in the mask by
$1-\exp\left(-9\theta^2/(2\theta_{\rm ap}^2)\right)$ with $\theta_{\rm
  ap}=15'$. All three maps agree very well. There is slightly larger noise
in LIL map which shows up as bigger error bar in the smallest scale bin.
The bin sizes were chosen to be large enough to minimize the
  correlation between the neighbouring bins, which are expected when we do
  not have full sky, and small enough so that there are enough bins that
 the shape of the power spectrum is not washed out.

  It is interesting to note there is a maximum at 1 Mpc ($\sim\,$$R_{500}$)
  and a decline, which corroborates the idea (\S3) that this scale is the
  dominant dynamical scale, even for absolute fluctuations -- such scale is
  commonly associated with major mergers and cosmological filamentary
  inflows. This interpretation comes with a
    caveat. On small scales we expect suppression of power because of the
    integration along the line of sight which suppresses small scale power
    in the 2-d projected map. On large scales we expect decrease in power
    because the outskirts of the cluster become important and there the
    amplitude of the average SZ signal declines. 
  In particular, the above mentioned effects would shift the peak in
  the projected power spectrum. The peak is on the same scale as the large
  fluctuation visible slightly off center in the $\delta y$ map in
  Fig. \ref{Fig:szfluc}.

In order to test that the fluctuations we are seeing are intrinsically from the
cluster and not SZ effect from background sources or contamination, we
randomly select three test regions near the Coma cluster, but at least few
degrees away from both Coma and Virgo clusters. 
Since there is no dominant
cluster in these regions, any fluctuations would be dominated by background
SZ signal and contamination. The regions chosen are at similarly high
galactic latitudes and so should have the similar magnitude of
contamination. The centers of the three test regions in galactic
coordinates $(l,b)$ are $(171.6,84.0)$, $(75,81.0)$ and $(315,86.0)$ and we
will label them as ``region 1/R 1'', ``region 2/R 2'', and ``region 3/R 3''
respectively, as shown in Fig. \ref{Fig:map}. 

We apply the same procedure that we used to calculate the
fluctuation power spectrum of Coma to these three regions. The results are
shown in Fig. \ref{Fig:testfluc} for the LIL  map. We expect that all
contamination and non-Coma signal in the Coma regions  would be similar to
that in the test regions. From our null test in Fig. \ref{Fig:testfluc} we see that our
power spectrum is dominated by the SZ anisotropies originating in Coma. 
The last bin is affected by contamination (at $1\sigma-2 \sigma$ level) giving a positive
  signal in the test regions; at the smallest scale, we thus expect
  some contribution from contamination, albeit not fully
  overwhelming the intrinsic Coma signal. 
  We note that
  this test just shows the typical contamination in this part of the sky is
small compared to our signal. The actual contamination in the Coma cluster
could be smaller since we expect the galactic contamination to be smaller,
or could be larger if there are additional radio and infrared point
sources in the Coma cluster that we cannot account for by looking at other
regions of the sky. The contribution of the point sources to the NILC,
MILCA and LIL maps is very different \citep{lily} with NILC probably the
cleanest in this respect. The fact that the results from the three maps
agree so well with each other gives us confidence that the contribution of
point sources 
and other extra-galactic and galactic contamination is negligible. Also, there are no strong point sources
detected in the region of interest in the Planck point source catalogs \citep{pccs2}.

\section{Pressure fluctuations in Coma cluster}
\begin{figure}
\resizebox{\hsize}{!}{\includegraphics{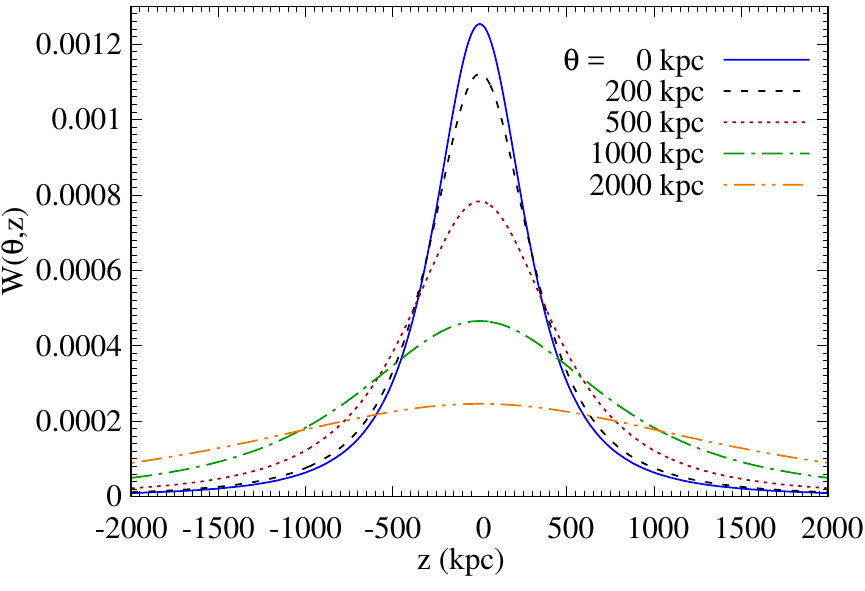}}
\caption{\label{Fig:window}The window function $W(\theta,z)$ at several
  values of the projected distance $\theta$ from the center. As we get away from the center, the window
  profile becomes broader increasing the suppression of the small scale power.}
\end{figure}
\begin{figure}
\resizebox{\hsize}{!}{\includegraphics{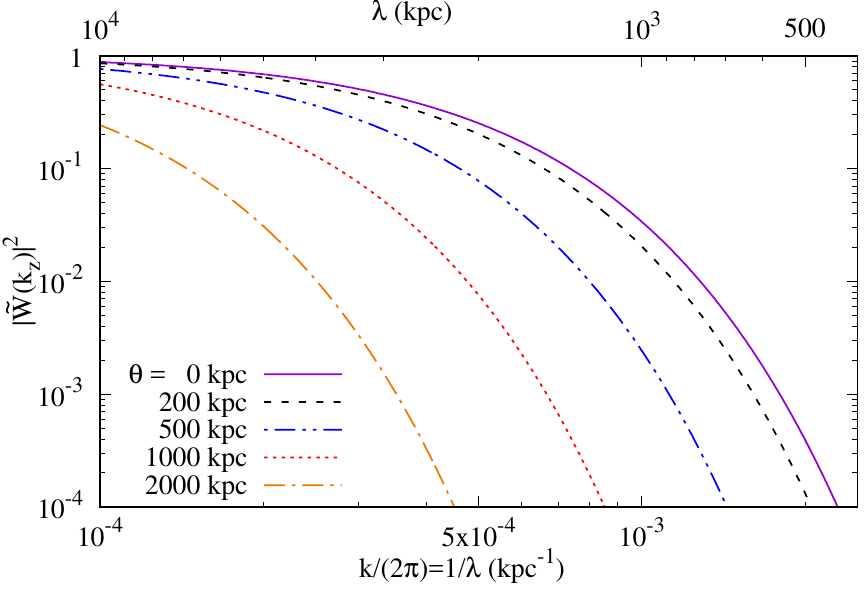}}
\caption{\label{Fig:win}The window function $|\tilde{W}(k_z)|^2$ at several
  values of the projected distance $\theta$ from the center. As we get away from the center, the window
  profile becomes broader increasing suppression of the small scale power.}
\end{figure}

We want to calculate the 3-d power spectrum of fractional pressure fluctuations
$\delta P/P$.
We can write the 2-d SZ power spectrum, $P_y$, as a convolution of the
3-d pressure power spectrum $P_P$ with a window function (Appendix \ref{appb} for a derivation; see also \citealt{Peacock:1999}) as 
\begin{align}
P_y(k_{\theta})=\int\frac{\id
  k_z}{2\pi}\left|\tilde{W}(k_z)\right|^2P_P\left(\left|\bkt+\bkz\right|\right),\label{Eq:2d3d}
\end{align}
where $P_y$ is the power spectrum of fractional $y$ fluctuations  $\delta y/\bar{y}$, $P_P$ is the power
spectrum of fractional pressure fluctuations, $\delta P/\bar{P}$, $\bkt$ is
the Fourier vector in the 2-d x-y plane of the sky and $\bkz$ is the
Fourier vector along the line of sight direction $\bhz$, non-bold quantities
are the scalar amplitudes of the corresponding vectors and $\tilde{W}(k_z)$ is the
Fourier transform in the $\bhz$ direction of the 
weight function defined by
\begin{align}
W(\bt,z)\equiv \frac{\me c^2}{\sigT} \frac{\bar{P}(\bt,z)}{\bar{y}(\bt)},
\end{align}
with $\bt=(\theta_x,\theta_y)$ is the position vector in the x-y plane of the sky in the flat
sky approximation so that its magnitude $\theta$ is the projected distance
from the center of the cluster. Note that the window function, plotted in
Fig.~\ref{Fig:window} in real space for several values of projected distance $\theta$ for the best
fit $\beta$-model profile, is in general
a function of position on the sky, $\bt$, which we have ignored in deriving
the equation \ref{Eq:2d3d}. This is a good approximation for the central part of the cluster. We show in
Fig. \ref{Fig:win} the Fourier space window function $|\tilde{W}(k_z)|^2$ at several
values of projected distance
$\theta$ from the center of Coma calculated using our best fit profiles in
Eqs. \ref{Eq:yprof} and \ref{Eq:pprof}. Assuming  $|\tilde{W}(k_z)|^2$ to be
independent of $\theta$ is therefore a rough approximation which becomes
more and more accurate as we confine ourselves to the smaller and smaller
projected region of the cluster.

\begin{figure}
\resizebox{\hsize}{!}{\includegraphics{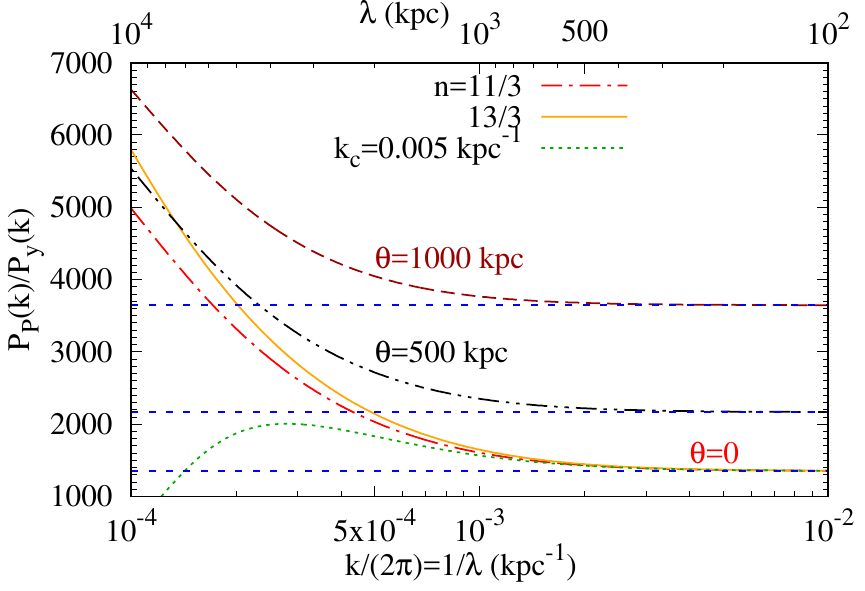}}
\caption{\label{Fig:2d3dratio} The ratio of pressure (3-d) to SZ (2-d) power
  spectra for different 3-d power spectra and projected radial distances $\theta$. The
  default values for all curves unless specified are $k_c=10^{-3}~{\rm
    kpc}^{-1}$ and $n=11/3$. The Eq.~\ref{Eq:2d3dapp} approximation 
  is shown as the horizontal line.}
\end{figure}

\begin{figure}
\resizebox{\hsize}{!}{\includegraphics{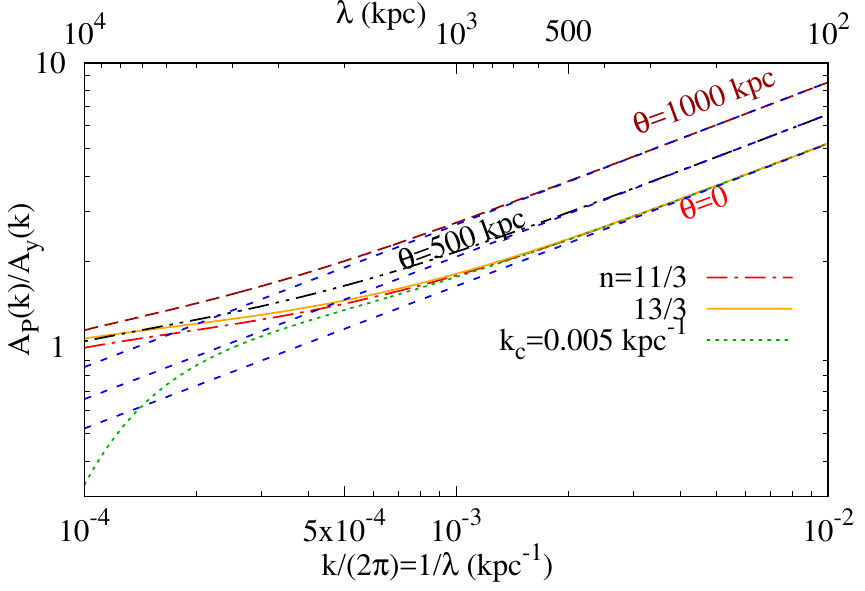}}
\caption{\label{Fig:a2d3dratio} The ratio of pressure (3-d) and SZ (2-d)
  fluctuations amplitude spectrum
  for different 3-d power-law slopes and projected distances $\theta$. The Eq.~\ref{Eq:2d3dapp} approximation 
  is shown as the straight lines. 
  Note that the amplitude ratio remains near unity at large scales, regardless of projection effects.
  For a comparison to the X-ray window see \citet{zc2012}.
  The green curve uses an aggressive exponential cutoff for illustrative purposes;
    more likely, the 3-d spectrum merges with the
    cosmological power spectrum on scales larger than the cluster size, which is beyond our reach.}
\end{figure}

We can further simplify Eq. \ref{Eq:2d3d} if we confine ourselves to small
scales so that $k_z\ll k_{\theta}$. 
We see from Fig. \ref{Fig:win} that the
window function drops very sharply on small scales and the contribution to
the convolution integral from $k_z/(2\pi)\gtrsim 5\times 10^{-4}~ \rm{kpc}^{-1}$ would be small
justifying this approximation for physical scales $\lesssim2000$\,Mpc.
In this limit, the relation between 2-d SZ
power spectrum and 3-d pressure power spectrum reduces to \citep{churazov2012}
\begin{align}
P_y(k_{\theta})&\approx P_P\left(k_{\theta}\right)\int\frac{\id
  k_z}{2\pi}\left|\tilde{W}(k_z,
\theta)\right|^2\nonumber\\
&\approx N({\theta})\; P_P\left(k_{\theta}\right)\label{Eq:2d3dapp}.
\end{align}
For $\theta=\{0,100,200,500,1000,2000\}~{\rm kpc}$ we get
$N=\{7.4,7.2,6.6,4.6,2.7,1.4\}\times 10^{-4}$ respectively.
We can check the last approximation explicitly by using an input 3-d power
spectrum and performing the convolution integral. For a power law power
spectrum  exponentially cut off at small and large scales,
$P_p(k)=k^{-n}\exp\left({-k_c/k}\right)\exp\left(-k\right)$, we
show the ratio of 3-d to 2-d power spectrum in Fig. \ref{Fig:2d3dratio} for
different parameters $n$ and $k_c$ and different projected distances. The
default values are $n=11/3$ and $k_c=10^{-3}~{\rm kpc}^{-1}$ unless
specified otherwise. 
The approximation in Eq. \ref{Eq:2d3dapp} is quite
accurate on small scales but starts to deteriorate on large scales where it is
sensitive to the exact form of the power spectrum, in particular the cutoff
scale $k_c$. Because of the information lost in projection it is impossible
to accurately deconvolve the power spectrum; we will thus use for the rest of the paper 
the approximation in Eq. \ref{Eq:2d3dapp} with $N=7\times 10^{-4}$.
Although in terms of power spectral density $P_k$ the ratio $N$ appears to be large, if we consider
the amplitude spectrum $A_k$ (Eq.~\ref{eq:Ak}) -- which is
  dimensionless having the same units as the corresponding real space perturbation and
  is the power per unit logarithmic $k$ interval -- the $A_P(k)/A_y(k)$ ratio remains near unity at large scales, increasing only by a factor of a few for scales smaller than 500 kpc (see Fig.~\ref{Fig:a2d3dratio}).

We now have all ingredients to calculate the power spectrum of SZ
fluctuations for Coma and use Eq. \ref{Eq:2d3dapp} 
to scale the 2-d SZ fluctuation power spectrum to 3-d pressure fluctuation power
spectrum. We first divide the SZ fluctuation map of
Fig. \ref{Fig:comayfluct} by the average profile to get a map of fractional
perturbations of SZ shown in Fig. \ref{Fig:comayfluctnormed}.
\begin{figure*}
%\subfigure{{\includegraphics[scale=1.3]{pk_coma_normed.pdf}}}
\subfigure{{\includegraphics[scale=0.335]{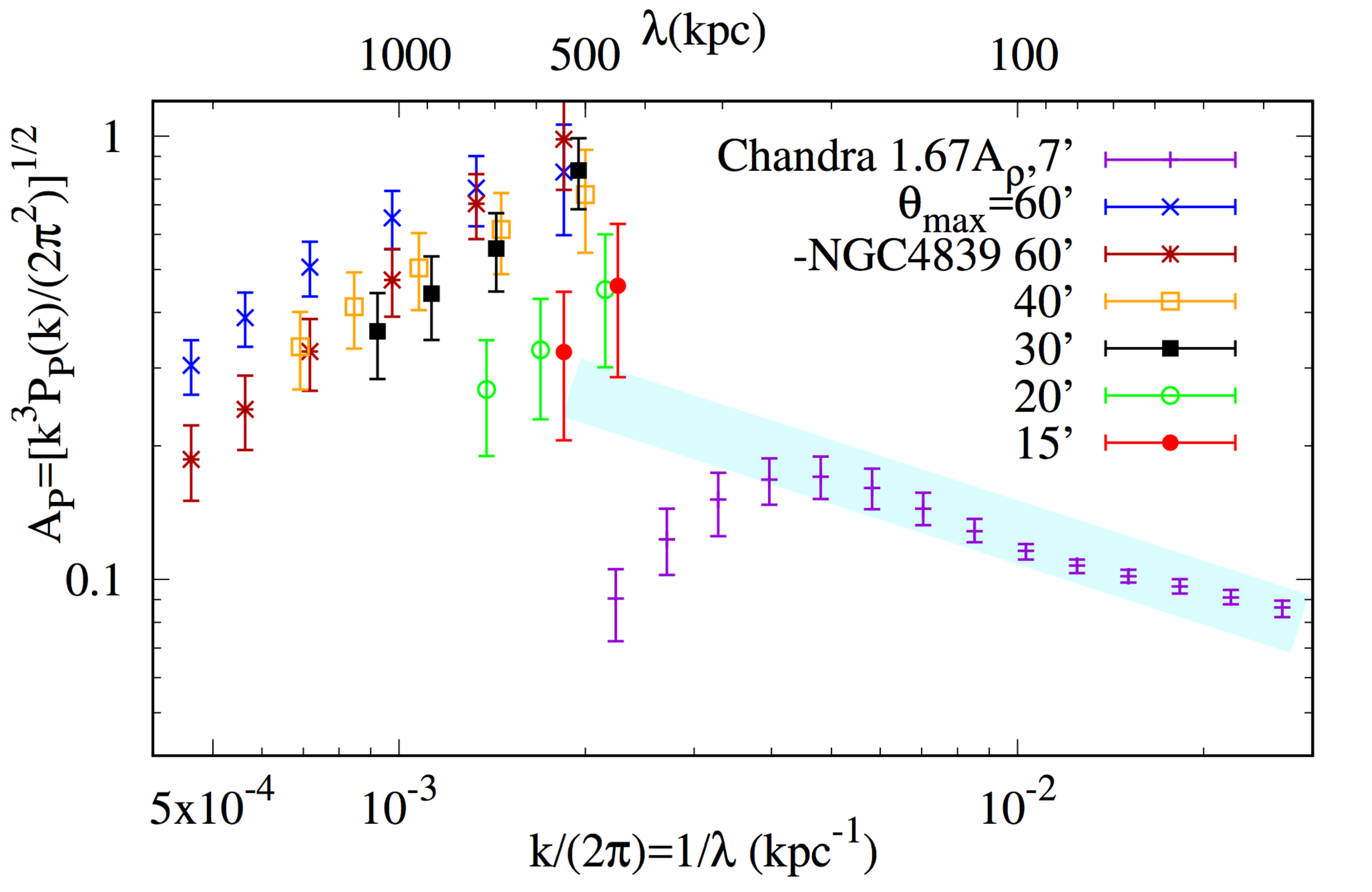}}}
\caption{\label{Fig:Pppower}Amplitude spectrum of pressure fluctuations for
  varying extraction radius $\theta_{\rm max}$ (LIL maps), for {\it Planck}
  and {\it Chandra} data (violet; \citealt{Gaspari:2013}). This key plot
  shows that large radii have significantly larger perturbations, and thus
  stronger turbulence motions (${\rm Ma_{3d}}\gtrsim0.8$), consistently
  with cosmological mergers and inflows. The injection scale is constrained
  to be at $\approx 500$ kpc, given the decline of amplitude at small
  $k$. The power spectrum labeled ``-NGC4839 60' '' is obtained by masking
  15' radius region with 15' apodization centered on the NGC 4839 group of
  galaxies. Note that the {\it Chandra} power spectrum is calculated in
    box of half-size 7' which is below the Planck resolution. The half-size
    of the box at the
    Coma distance is $\sim 200~ \rm{kpc}$ which coincides with the peak in
    the X-ray derived power spectrum. The SZ and
    X-ray data thus probe mutually exclusive different regions of the cluster (outer and inner part, respectively).
    Extrapolating the {\it Chandra} inertial range (cyan band) shows
    the 2 datasets are consistent within errors in the inner extraction
    region. 
}
\end{figure*}
Since the profile $\bar{y}\rightarrow 0$ at large
distances from the center, the fractional fluctuations, $\delta y/\bar{y}$
would diverge and far from the cluster the noise and contamination would
start getting amplified and will dominate the overall power spectrum.  The power spectrum is therefore
a strong function of the maximum radius we decide to use for 
calculating the power spectrum. We show in Fig. \ref{Fig:Pppower} the
power spectrum of pressure fluctuations, $\delta P/\bar{P}$, for different
values of maximum radius from the center of Coma, $\theta_{\rm max}$,
i.e. we mask the regions beyond the maximum radius in the map leaving only
the region $\theta< \theta_{\rm max}$ unmasked. We further apodize
the mask with $\theta_{\rm ap}=15'$ Gaussian 
as defined in the previous section for
$\theta_{\rm max}>30'$ and use $\theta_{\rm ap}=5'$ for $\theta_{\rm max}
\le 30'$\footnote{Apodization is beneficial for reducing the edge effects while calculating the power spectrum, but 
    it should not be so large that most of the signal is obscured. Large
    apodization implies using less data, hence larger error bars.}.
We note that formally we retrieve electron pressure via SZ signal, however, since we are only interested in relative fluctuations, the ion component is secondary here. Moreover, \citealt{Gaspari:2013} found 
that the ion and electron temperature differ only by 1\,-\,15 percent (core
to outskirts, respectively) in typical ICM turbulence
simulations. Note that the errors on large scales are dominated by
  the sample variance. The sample variance is a function of bin size and
    for large bin size the variance does not rise as fast on large scales as we would expect
    for unbinned or logarithmically binned data since we are
    averaging many modes even on large scales.

\begin{figure}
\centering
\vspace{-0.1cm}
\subfigure{{\includegraphics[scale=0.89]{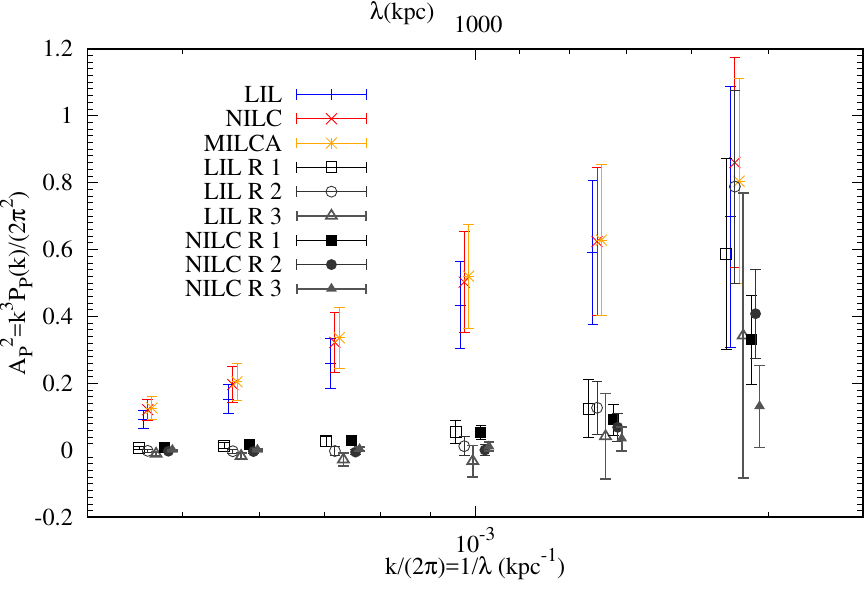}}}
\subfigure{{\includegraphics[scale=0.89]{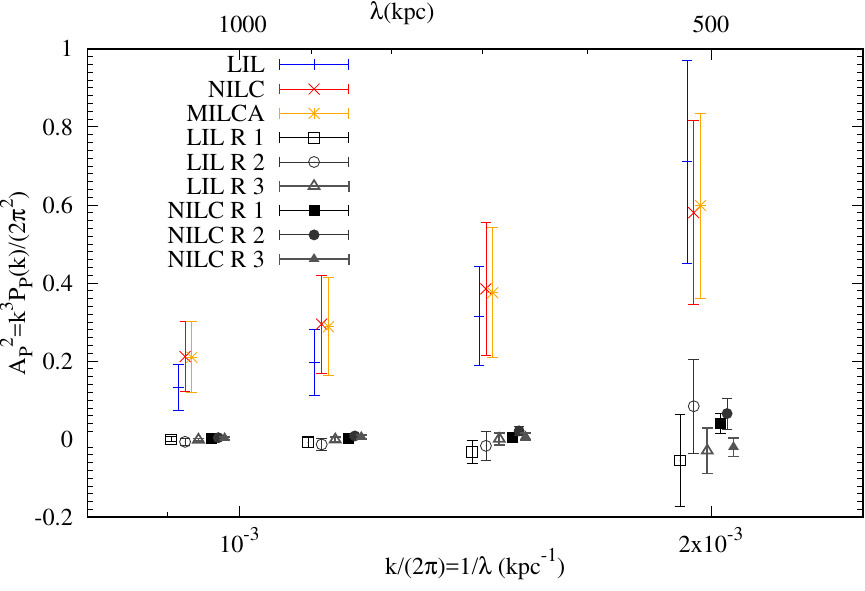}}}
\subfigure{{\includegraphics[scale=0.92]{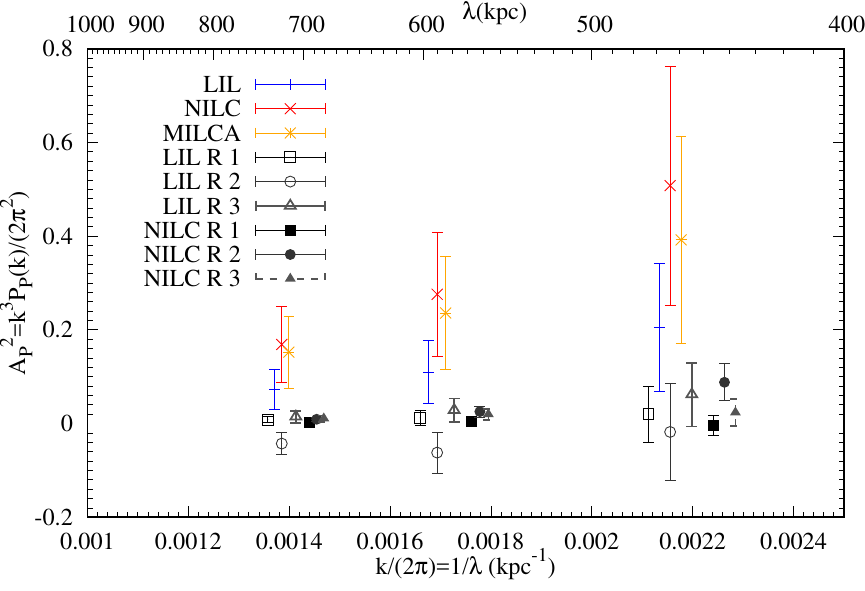}}}
\vspace{-0.3cm}
\subfigure{{\includegraphics[scale=0.92]{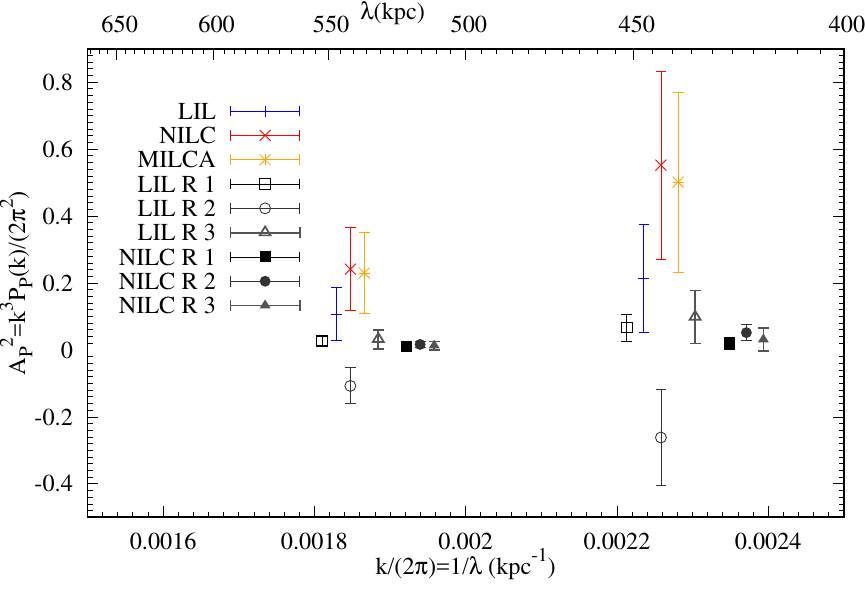}}}
\caption{\label{Fig:test60} Squared amplitude spectrum of Coma pressure fluctuations for
  $\theta_{\rm max}=60',30',20',15'$ (top to bottom) in LIL, MILCA and NILC maps and in the
  test regions (LIL maps). For the test region we used our best fit
  Coma profile in the denominator to calculate $\delta y/\bar{y}$.}
\end{figure}

As expected, the fluctuations when normalized to the average profile
$\bar{P}$ are larger when we use a larger extraction radius. 
For comparison, we show
the amplitude of density fluctuations retrieved from the latest {\it Chandra} X-ray observation (\citealt{Gaspari:2013}; 
updated version of the data shown in
\citealt{churazov2012})\footnote{{As in \citet{Gaspari:2013}, we
    prefer to compare only with the more robust and improved {\it Chandra}
    data, as {\it XMM} has 40$\times$ lower on axis angular resolution. 
    There are, in addition, systematic differences between Chandra and XMM
    instrument  \citep[e.g.][]{srl2015} }}. 
Our smallest extraction radius of 15' is larger than 
 the half-size of the X-ray extraction region of 7'. This fact combined
 with our SZ map resolution of 10' means we are 
 probing different regions of the Coma cluster in SZ and X-rays. In the
smallest disc  we have also non-negligible contribution from
contamination. The amplitude is however still consistent with $\gamma\,
A_{\rho}$ within 2-$\sigma$, where $A_\rho$ is the 3-d amplitude of power
spectrum of density fluctuations $\delta \rho/\bar{\rho}$ from {\it
  Chandra}, and $\gamma=5/3$ is the adiabatic index (\S3.1 for the
discussion). We
point out that the decrease in the large scale power in X-ray data
is an artifact of approaching the limited box size. This becomes clear
when we compare the {\it Chandra} and {\it XMM} data in \citet{churazov2012} (their Fig.~13).
{The {\it XMM} box size is 12' and the downturn in power spectrum happens at
larger scales in the {\it XMM} data compared with the {\it Chandra} data}. There is also
discrepancy on large scales between 
the {\it XMM} and {\it Chandra} pointing to systematic effects related to the box size
(see Fig.~14 in \citealt{churazov2012}).
{E.g., above 300 kpc, \citeauthor{churazov2012} (2012; section 5.1) 
claim a factor of 2 uncertainty in systematic errors.}
If we extrapolate our amplitude to the core size of 7' probed
by X-ray observations, the agreement between the X-ray and SZ improves.
Similarly, if we extrapolate the X-ray cascade to several 100 kpc scale, the inertial range joins the
SZ green and red points within errorbars (cyan band).
%are consistent with the X-ray data.
Keeping in mind the discussed uncertainties
and the fact we are probing different extraction regions of the cluster, 
the SZ results are consistent with the X-ray data.

{The 60' radius region includes the group of galaxies NGC 4839
  which is falling into the Coma cluster \citep{briel1992,neumann2001}. This group is marked
  with a white circle in Fig. \ref{Fig:comayfluctnormed} and shows up as a
  bright spot in the normalized fluctuations. %shown in the figure. 
  To estimate the contribution of this group to the fluctuation signal, we
  mask 15' radius region around the hot spot (which coincides with the
  X-ray source in ROSAT all sky survey; \citealt{briel1992}). In addition we apodize
  the combined mask for the 60' region analysis with 15' Gaussian as
  explained above, down-weighting the contribution from extended 15'-30' radius region
  around the group. Comparison of the two 60' power spectra -- with and
  without masking the group -- shows that the contribution from
  the group is small, decreasing the amplitude at large scales by 20\,-\,30 percent. We note
  that the group lies outside the 30' radius and, taking into account the
  fact that we extend our masks by apodization, it does not contribute to the
  fluctuations calculated in the 40' and smaller radius regions.
  It should also be noted that the hot plasma is volume filling within the cluster and internal groups. So
  cutting regions of the sky will inevitably also remove some  turbulence
  perturbations intrinsic to Coma. 
}

To test whether we are seeing actual anisotropies from Coma or the
contamination from non-Coma background SZ or foregrounds, we again use our test
regions. We compare the test regions in LIL and NILC maps with the Coma region in LIL,
NILC and MILCA maps in Fig.~\ref{Fig:test60}
for $\theta_{\rm max}=60', 30', 20',15'$. We plot $A_k^2$ in
these plots since when we are dominated by noise in the test regions the
cross spectrum between the two half-ring maps can take negative values.
We see that in Fig. \ref{Fig:test60} (top) we can attribute all signal measured
in the Coma region to the anisotropies in the Coma cluster except in the
last bin where the contamination may have a sub-dominant contribution.
Note that as we take the cross-spectrum, the noise cancels out in the mean signal but still contributes to the error bars, which are thus increasing toward the resolution limit.
For our smallest disc with radius $\theta_{\rm max}=15'$ (Fig. \ref{Fig:test60}, bottom) we
can measure only the smallest scales and here our measured power in Coma is of
similar magnitude as the test regions. The NILC map test regions have
however smaller contamination, while the signal in NILC is consistent with
that in the LIL and MILCA maps. We therefore have evidence of pressure
fluctuations even in the central part of the Coma which cannot be fully
explained by contribution due to contamination.

\begin{figure}
\resizebox{\hsize}{!}{\includegraphics{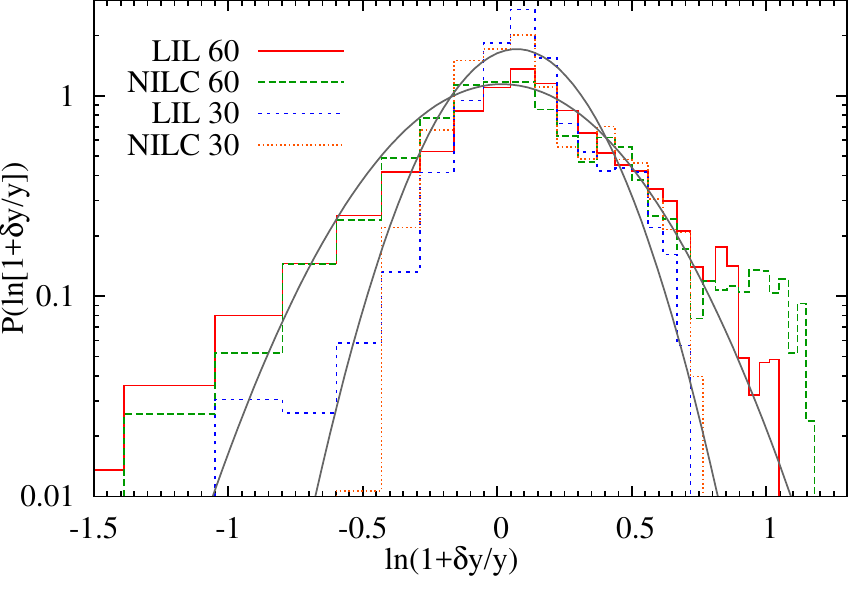}}
\caption{\label{Fig:pdf} 
{The PDF of SZ fluctuations in the Coma cluster within the $60'$ and $30'$ radius
of the center in NILC and LIL maps. The distribution of $\ln(1+\delta
y/\bar{y})$ is consistent with a Gaussian distribution (shown by solid gray
lines) and therefore $\delta y/\bar{y}$ is consistent with having a
log-normal distribution with mild non-Gaussianities (in $\ln(1+\delta
y/\bar{y})$), which is also the statistics of 
thermodynamic perturbations driven by turbulence.}
}
\end{figure}

We plot the probability density function (PDF) of $\ln\left(1+\delta
  y/\bar{y}\right)$ in Fig.~\ref{Fig:pdf} and the best fit normal
  distributions for $60'$ and $30'$ regions. 
  The PDF of $\delta y/\bar{y}$
  is approximately consistent with being a
  log-normal distribution. The properties of the distribution of
  $\ln(1+\delta y/\bar{y})$ (mean, variance, skewness, and kurtosis) 
  are given in Table \ref{Tbl:pdf} for all three maps. 
  The results from the three
  maps are again consistent with each other, especially for the mean and
  variance indicating that the contamination is sub-dominant.
  The spectra and related PDF have key implications for the ICM physics, which we discuss in depth in the next section.

\begin{table}
\caption{\label{Tbl:pdf}Statistical properties of distribution of
  $\ln(1+\delta y/\bar{y})$ in the Coma cluster within radius of $60'$ and $30'$.}
\begin{tabular}{|c|c|c|c|c|}
\hline 
& Mean & Variance & Skewness & Kurtosis excess\\
\hline
LIL $60'$&$-0.054 $ & $ 0.26 $ & $ -2.1 $ & $12.3 $\\
MILCA $60'$&$-0.1 $ & $ 0.24 $ & $ -1.6 $ & $13.5 $\\
NILC $60'$&$-0.035 $ & $ 0.20 $ & $ -1.1 $ & $11.3 $\\

LIL $30'$&$0.033 $ & $ 0.062 $ & $ -0.73 $ & $3.23 $\\
MILCA $30'$&$-0.034 $ & $ 0.066 $ & $ 0.19 $ & $0.07 $\\
NILC $30'$&$0.035 $ & $ 0.060 $ & $ 0.54 $ & $-0.09 $\\
\hline
\end{tabular}
\end{table}

\subsection{Systematics due to ellipticity of the cluster}
\begin{figure*}%[H]
\subfigure{\includegraphics[scale=0.23]{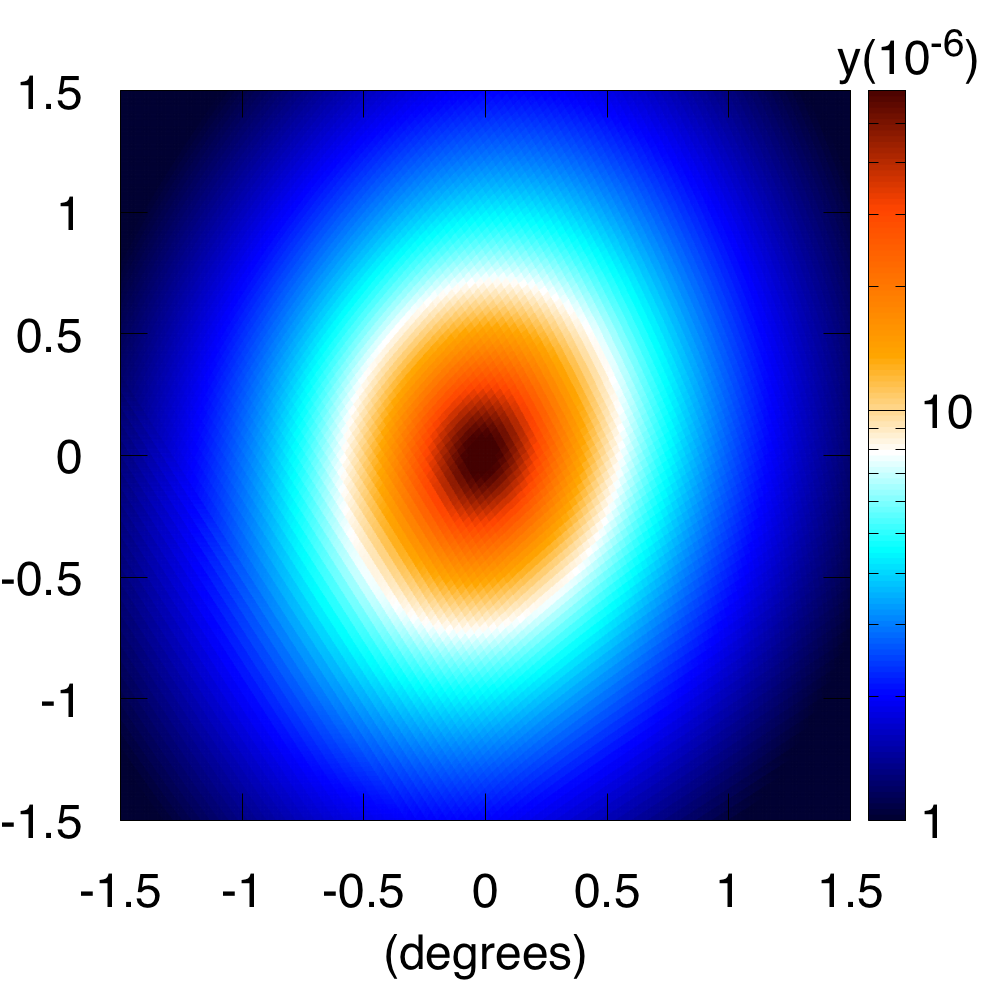}}
\subfigure{\includegraphics[scale=0.23]{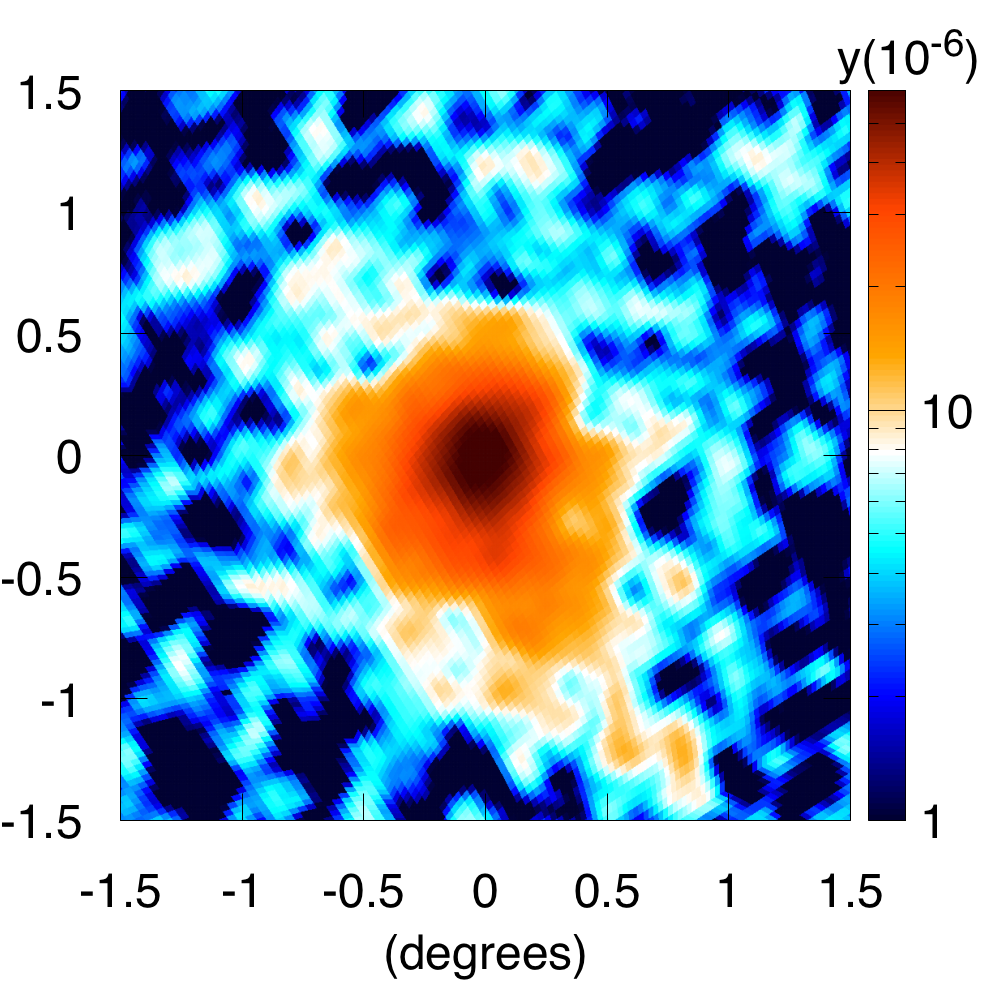}}
\caption{\label{Fig:comaellip} 
The elliptical cluster model (left) and Coma cluster in LIL map (right). We remark
Coma cluster is less elliptical than our model with eccentricity
$e=0.6$. Therefore, the contribution to the power spectrum from ellipticity obtained from such model
shall be seen as upper limits; in reality, we expect the ellipticity contributions to be much smaller.
}

\subfigure{\includegraphics[scale=0.225]{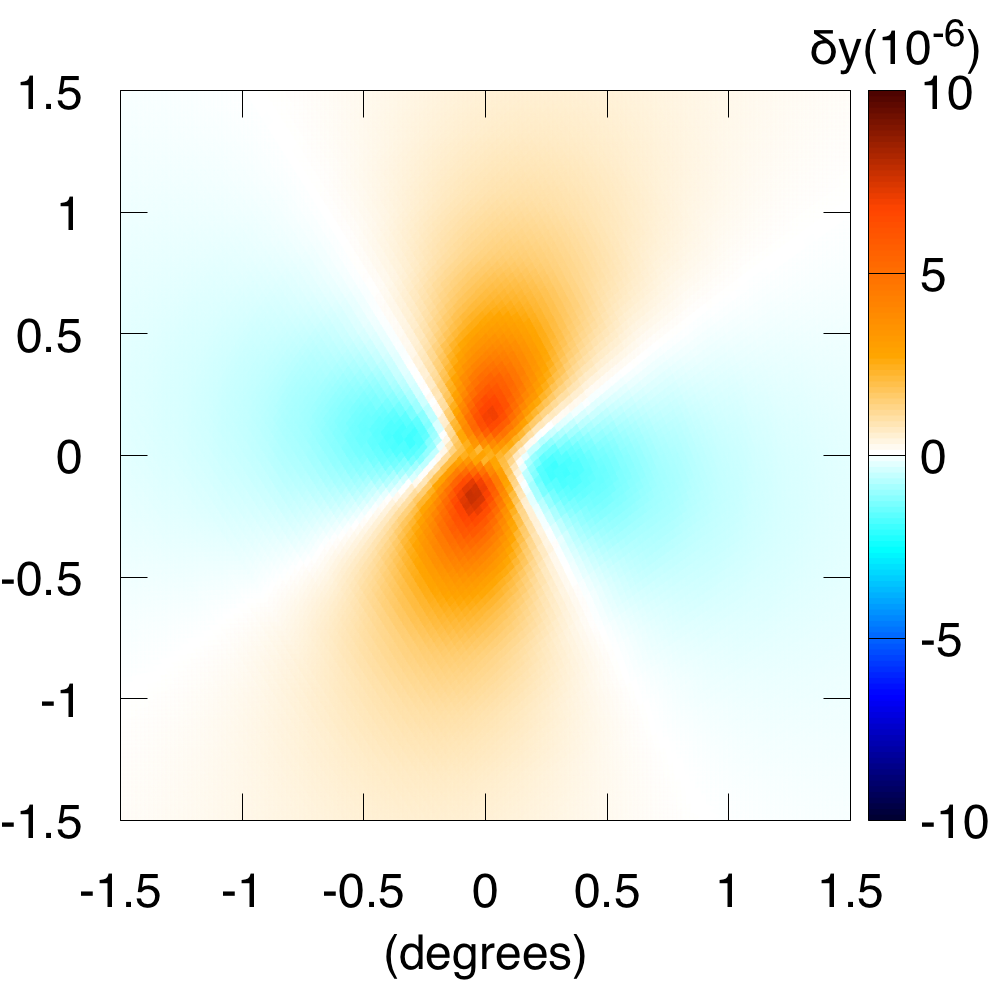}}
\subfigure{\includegraphics[scale=0.225]{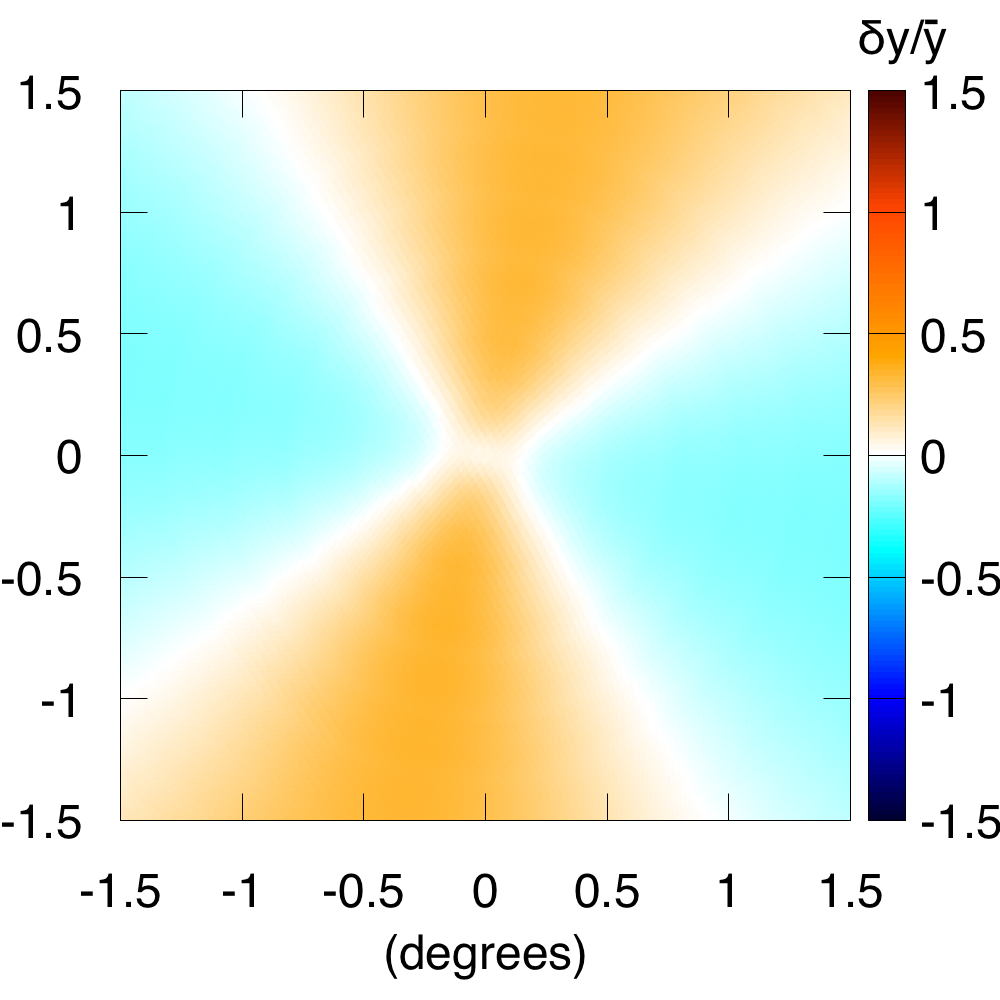}}
\caption{\label{Fig:ellipdy} 
SZ fluctuations (left) and normalized fluctuations (right) for the
elliptical cluster model. These maps should be compared with the
corresponding Coma maps in Figs. \ref{Fig:comayfluct} and \ref{Fig:comayfluctnormed}.
}
\end{figure*}

\begin{figure}
\resizebox{\hsize}{!}{\includegraphics{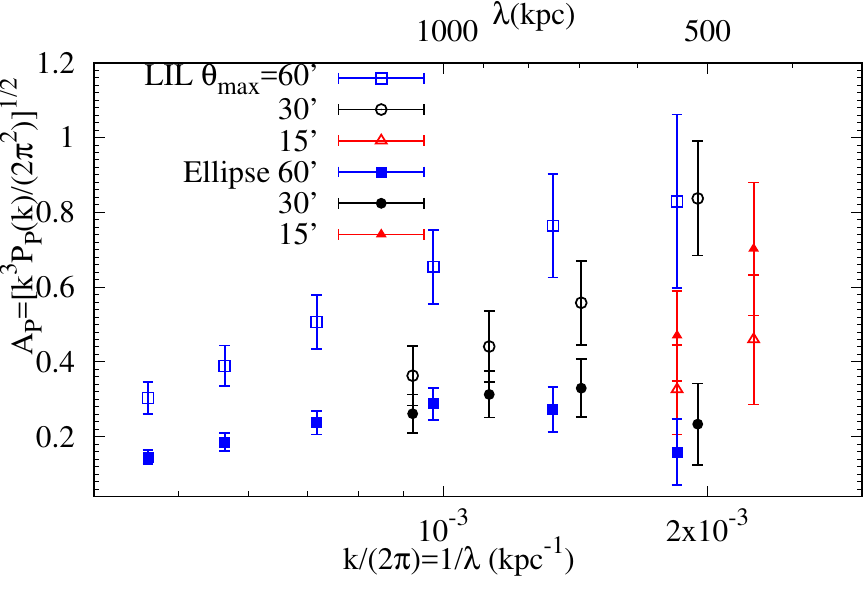}}
\caption{\label{Fig:ellipcomp} 
The amplitude of the power spectrum of the elliptical model compared with the
measured fluctuations in Coma cluster.
}
\end{figure}

We have so far assumed spherical symmetry in our analysis. Clusters in the
standard cosmological model are expected to be elliptical and this is
indeed true for the Coma cluster \citep{sk1978,comaellip}. If we try to fit a spherically or
circularly symmetric profile to the ellipsoidal data, the residuals will
look like anisotropies and may contribute to the fluctuation power
spectrum. To study the amplitude of this effect we make an artificial
elliptical cluster with similar $\beta$-profile as the Coma cluster and
repeat all our steps on this simulated cluster. \citet{comaellip} have
tried to fit an elliptical profile to the XMM-Newton observations getting
eccentricities for the core of the cluster of $e=0.4,0.6$ for the pn and MOS data,
respectively. We use the higher eccentricity as reference for our elliptical
cluster model. We show the elliptical model and Coma cluster in the LIL map in
Fig.~\ref{Fig:comaellip}.
We note Coma cluster gas distribution is much more complex than a simple elliptical
profile and different radii have elongations along different
directions. We then perform our previous analysis on such elliptical
cluster model, subtracting the azimuthally averaged profile and retrieving $\delta y$
and $\delta y/\bar{y}$ maps. These maps are shown in Fig. \ref{Fig:ellipdy}
and should be compared with the corresponding Coma maps in
Figs. \ref{Fig:comayfluct} and \ref{Fig:comayfluctnormed}. The level of
anisotropy introduced by ignoring the ellipticity is negligible compared with
the actual fluctuations observed in the Coma cluster. 
In Fig.~\ref{Fig:ellipcomp}, we show the amplitude of
the power spectrum of normalized fluctuations for the elliptical model cluster and
compare it with the fluctuations in the Coma cluster in LIL map. Except for the innermost core, the contribution
of ellipticity to the power spectrum is small (60' and 30' region) and can
be ignored on small scales. For
the core 15' region we are at the limit of resolution for the $y$ map; there
are indeed not enough pixels to make a definite statement about the
ellipticity. 
In passing, we note that subtracting a spherical profile from an elliptical
profile would result in a characteristic quadrupolar anisotropy which we
do not see in the real Coma map.
To summarize, the above results justify the usage of spherical symmetry and azimuthal
averages in our fiducial analysis.
%\newpage
\section{ICM physics: turbulence, thermodynamics, HSE bias, and cosmological implications}

As introduced in \S1, the dominant scale and amplitude of the power spectrum allow us to probe the physics and (thermo)dynamics of the ICM plasma.
The retrieved spectrum of pressure
fluctuations at 500 kpc scale has an amplitude of $33\pm 12$\% in the inner 
$15'$ extraction region of the cluster. The amplitude of pressure
fluctuations at the same scale increases to $74\pm 19\%$ if we use the $40'$ radius region.
Using previous Eq.~\ref{Ma_rho} -- {which arises from the simulation modeling carried out in \citet{Gaspari:2013}} -- and the adiabatic mode ($\delta P/P\simeq\gamma\,\delta \rho/\rho$; discussed below), we estimate that turbulence is characterized by 3-d Mach number 
${\rm Ma_{3d}} = 0.8\pm0.3$ (or ${\rm Ma_{1d}} \approx 0.5\pm0.2$).
This is larger value than the {\it Chandra} estimate 
since the new observational constraint on the amplitude peak has increased by $2\times$.\footnote{
We note that the injection scale can not be derived from the \citet{Gaspari:2013} modeling, but 
must come from the observational data.}
As the adiabatic sound speed is $c_{\rm s}\simeq1.5\times10^4\,T^{1/2}\simeq1.5\times10^3$\,km\,s$^{-1}$ ($T_0=8.5$\,keV), the large-scale eddies have a characteristic velocity $\sigma_v\approx1.2\times10^3$\,km\,s$^{-1}$.
For the $40'$ region, we find ${\rm Ma_{3d}}=1.8\pm 0.5$ if we extrapolate
Eq.~\ref{Ma_rho} to the supersonic regime, which is currently untested. We
note that the temperature decreases down to 2 keV at 2 Mpc radius in Coma
(\citealt{Simionescu:2013}), therefore the physical velocities do not
double as the Mach number does but increase by a much smaller amount, $\sigma_v\approx1.8\times10^{3}$, while the compressibility of the medium progressively increases. The fact that at larger radial annuli the ICM reaches the transonic value 
is likely an indication that we are approaching the accretion shock region.
Large-scale velocities up to a few $10^3$\,km\,s$^{-1}$ are consistent with {\it Chandra} constraints derived from the redshift maps of 6 massive clusters undergoing significant merging activity (\citealt{Liu:2016}), as well as large-scale cosmological simulations (e.g., \citealt{Schmidt:2016}).

At scales larger than 500 kpc, the pressure fluctuations in all spectra
(corresponding to different maximum extraction radii\footnote{Note that the scale
  $\lambda$ (or Fourier mode $k$) is different from the radial distance
  $r$, the latter is the region used for calculating the power
    spectra while the former describes the wavelength of fluctuations.}) steadily decline, indicating a dominant turbulence injection scale at that location.
The injection $L_{\rm inj}\simeq500$\,kpc is consistent with gas motions driven by mergers and large-scale inflows, as indicated by cosmological simulations (e.g., \citealt{Lau:2009,Vazza:2009, Vazza:2011}; \citealt{Miniati:2014})\footnote{{Nowadays it is still numerically unfeasible to resolve the long-term and 10s Mpc cosmological evolution, at the same time resolving turbulence and plasma physics down to the kpc scale over the whole cluster volume.}}. 
Even for the larger extraction region, the fluctuations power is mostly contained within $r < 1\,$Mpc, i.e., well within the virial radius.
Internal galaxy group merging and infall (e.g., \citealt{Eckert:2014}; \citealt{DeGrandi:2016})
is thus the likely dominant driver compared to the large-scale cosmological filamentary accretion.
Perturbations driven by the active galactic nucleus (AGN) feedback play a minor role, 
since the major AGN outflow outbursts can reach at best $0.1\,R_{500}$ (\citealt{Gaspari:2014_AGN}). 
The same can be said for perturbations linked to thermal instability
(\citealt{Gaspari:2015_TI}); besides Coma long cooling time, 
fluctuations related to thermal instability are typically contained within $r<50$\,kpc.
{External clumps accreted in the cluster outskirts may enhance density perturbations; however,
recent cosmological simulations have shown that such `clumpiness' is secondary compared with the perturbations
imparted by non-thermal gas motions (\citealt{Avestruz:2016}).
For instance, the SW large-scale perturbation is cospatial with an infalling sub-structure (\citealt{neumann2001}) 
%(e.g., \citealt{Eckert:2007}), 
which might in part increase the amplitude in the 60' spectrum; the 40' spectrum, on the other hand, excludes such region, showing a minor decrease in amplitude ($<20$ percent; Sec.~3) and thus signaling that most of the power is contained in the chaotic motions.}

Given the limited resolution of {\it Planck}, we are currently not able to constrain the small-scale (`inertial') slope of the power spectrum. The complementary X-ray data shows
a quasi Kolmogorov power-law ($P_k\propto k^{-11/3}$, i.e., $A_k\propto
k^{-1/3}$), classically related to turbulence with very low thermal conductivity. 
If we visually extrapolate the cascade beyond the {\it
  Chandra} peak (Fig.~\ref{Fig:Pppower}; cyan band), this joins within errorbars the {\it Planck} red/green data of the inner extraction regions (Fig.~\ref{Fig:Pppower}),
in line with a self-similar cascade of turbulent eddies from $L_{\rm
  inj}$ to the dissipation scale ($<30$ kpc).
  As warned in \citet{Gaspari:2013}, we remark the X-ray amplitude spectrum below $k<5\times10^{-3}$\,kpc$^{-1}$ is highly uncertain and can underestimate the perturbations, given the strong dependence of X-ray emissivity on radial distance ($\propto \rho^2$). A jump relative to SZ data is thus to be expected. %, given the different sensitivity between $y$ and X-ray signal. 
  Moreover, recall that {\it Planck} black/yellow/blue points cover much larger, more active extraction regions than that of {\it Chandra} (see Sec.~3).
  Higher resolution SZ observations are required to ensure tighter constraints on the full ICM
power spectrum and to directly link the SZ observations to the X-ray data and related ICM microphysics.
{\it Planck} data still add key value to our knowledge of the ICM physics,  
as it allows us to assess the global strength of internal motions and the ICM equation of state.
%(the isobaric versus adiabatic mode discussed below).

The distribution of SZ/pressure fluctuations follows a log-normal distribution with mild non-Gaussianities (fat tails and slight negative skewness; Fig.~\ref{Fig:pdf}).
Remarkably, turbulent motions drive thermodynamic perturbations (e.g., density, pressure) with similar log-normal statistics
(\citealt{Federrath:2010,Gaspari:2014,Porter:2015}). Log-normal fluctuations in the ICM have been also found by X-ray studies on smaller scales (\citealt{Kawahara:2008}). 
Regarding asymmetries, since the retrieved Coma Mach number is nearly transonic, a prediction from turbulence simulations (e.g., \citealt{Kowal:2007}) is that stronger compressive motions drive heavier tails, as found here. 
Overall, key signature of turbulence is to linearly drive the fluctuations rms $\sigma\propto {\rm Ma_{\rm 3d}}$ with log-normal and leptokurtic PDF, as we find in {\it Planck} data. 
This corroborates the importance of turbulence in shaping the dynamics of hot plasmas from the cosmological Mpc scale down to the kpc scales, and possibly further -- below the Coulomb collisional scale -- via kinetic plasma instabilities and Alfv\'en waves.

The {\it Planck} data helps us to constrain the dominant thermodynamic mode of the ICM.
Since the retrieved gas motions are dynamically important (${\rm Ma_{3d}}\gtrsim 0.8$),
perturbations are expected to drift toward the adiabatic regime (\citealt{Gaspari:2014}), i.e., $\delta P/P = \gamma\, \delta \rho/\rho$, the regime adopted in Fig.~\ref{Fig:Pppower}. This is valid for ${\rm Ma_{3d}>0.5}$.
The fact that we see substantial pressure fluctuations indicates by itself
that Coma cluster is globally not in isobaric mode, at least at intermediate and
large scales. In such regime, sound waves (linked to pressure fluctuations)
start to overcome pure entropy perturbations (mediated by buoyancy/gravity
waves). The latter regime would dominate if ${\rm Ma_{3d}\ll0.5}$ with
$\delta K/K \gg \delta P/P$, where $\delta K/K$ are the entropy perturbations (see \citealt{Gaspari:2014} for an in-depth theoretical derivation).
This is corroborated by the aforementioned cosmological simulations, 
showing that the cosmological accretion inflow and mergers drive turbulence with average kinetic energy $5$\,-\,$35\%$ of the thermal energy, from relaxed to unrelaxed clusters (e.g., \citealt{Vazza:2011}). 
Using our turbulence constraint, Coma has turbulent energy density in the inner region
\begin{align}
E_{\rm turb} \equiv \frac{1}{2}\,\rho\sigma_v^2= \frac{1}{2}\,\gamma (\gamma-1)\, {\rm
  Ma^2_{3d}}\; E_{\rm th}\simeq0.34\pm 0.25\,E_{\rm th}.
  \label{Eq:Eturb}
\end{align}
If we consider the 1 Mpc radius region, the turbulent energy becomes comparable to the thermal energy, although
it should be kept in mind the large error bars.
As expected, Coma belongs to the class of unrelaxed clusters with a major dynamic component.

Besides difficult lensing estimates, cluster masses are typically retrieved
via the X-ray density and temperature radial profile assuming hydrostatic
equilibrium (HSE), $\id P_{\rm tot}/\id r = -\rho\,GM(<r)/r^2$, where $G$ is
the gravitational constant, $P_{\rm tot}$ is the total pressure -- thermal
and non-thermal -- and $M(<r)$ is the mass enclosed within radius $r$.
Expanding the previous equation, the total mass can be retrieved as
\begin{equation}\label{e:HSE}
M_{\rm tot}(< r)= -\frac{k_{\rm B}T(r)\,r}{\mu m_{\rm p}\,G}\left[\frac{\id\ln\rho(r)}{\id\ln r} + \frac{\id\ln T(r)}{\id\ln r} +
\frac{P_{\rm nt}}{P}\frac{\id\ln P_{\rm nt}}{\id\ln r} \right],
\end{equation}
where the last term is the non-thermal pressure support which is not
included in observed X-ray masses $M_{\rm x}$, $\mu$ is the mean atomic
weight and $m_p$ is the proton mass. The mass bias is thus
\begin{align}     \label{e:bias}
b_{\rm M}\equiv \frac{M_{\rm x}}{M_{\rm tot}} -1 = -\frac{P_{\rm nt}}{P} \frac{\id \ln P_{\rm nt}}{\id \ln P} 
    \left(1+\frac{P_{\rm nt}}{P}\frac{\id\ln P_{\rm nt}}{\id \ln P}\right)^{-1},
\end{align}
Since $P=(\gamma-1)\,E_{\rm th}=(2/3)\,E_{\rm th}$ and $P_{\rm nt}=(1/3)\rho\sigma_v^2=(2/3)E_{\rm nt}$ 
(e.g., \citealt{Lau:2009, Schmidt:2016}), thus $P_{\rm nt}/P=(\gamma/3)\,{\rm Ma_{3d}^2}$
and 
\begin{align}     \label{e:bias2}
\frac{\id\ln P_{\rm nt}/\id\ln r}{\id\ln P/\id\ln r}= 1+2\,\frac{\id\ln{\rm Ma_{3d}/\id \ln r}}{\id\ln P/\id \ln r}.
\end{align}
Beyond the core radius, the pressure gradient slope is $-2\,(\beta+1/2)\simeq-3.14$ (Eq.~\ref{Eq:pprof}) and 
our retrieved Mach between the 40' and 15' radius implies $\id\ln {\rm Ma_{3d}}/\id\ln r\simeq0.8$, yielding a mass bias
\begin{align}     \label{e:bias3}
b_{\rm M}\simeq -0.27\,{\rm Ma_{3d}}^2\,\left(1+0.27\,{\rm Ma_{\rm 3d}^2}\right)^{-1}\simeq -15\% \div -45\%,
\end{align}
for the inner (15') and outskirt (40') region, respectively.
{The substantial increase in non-equilibrium measures at larger radii is consistent with large-scale cosmological simulations (e.g., \citealt{Battaglia:2015})}.
The above analysis accounts only for turbulence as the dominant source of non-thermal pressure; magnetic fields and cosmic rays may further increase the bias. \citet{Bonafede:2010} find a weak 1\,-\,5 $\mu$G magnetic field in Coma cluster (from the outskirts to the core, respectively) by using Faraday rotation measure in radio data. The magnetic pressure, $B^2/(8\pi)$, is thus $<10$ percent of our retrieved turbulent pressure.

Although we analyzed only one cluster, the HSE bias is expected to play a crucial role in the mass estimates of all cosmological systems, from massive clusters to galaxy groups (e.g., \citealt{Rasia:2004}; 
\citealt{Lau:2009,Sun:2009}).
Moreover, it is well known that cosmological parameter constraints (such as
$\sigma_8$) derived from SZ clusters and CMB are in tension \citep{cosmocluster2015}.
For instance, \citet{Shaw:2010} show that increasing the fraction of
$P_{\rm nt}$ support in the ICM reduces the total SZ power and can thus
alleviate such tension (see also \citealt{sptclusters2016}). Furthermore,
retrieving true masses allows us to accurately assess the main scaling
relations of virialized structures (in terms of both slope and intrinsic
scatter), such as $L_{\rm x}-M_{\rm tot}$ and $M_{\rm tot}-T_{\rm x}$,
where $L_{\rm x}$ is the X-ray luminosity and $T_{\rm x}$ is the X-ray temperature, typically within $R_{500}$
 (\citealt{Giodini:2013} for a review).
In conclusion, while total SZ power probes the thermal energy content, the related SZ fluctuations tell us the strength of the non-thermal deviations, thus providing a global, self-consistent view of the cluster (thermo)dynamics.

\section{Conclusions}
We have for the first time used the thermal Sunyaev-Zeldovich effect to constrain
the absolute and relative $y$ fluctuations. Such fluctuations put key constraints on the physics of the intracluster medium,
such as turbulence and the dominant thermodynamics, with major implications for cosmological studies.
We applied our methodology to the archetypal massive galaxy cluster Coma (Abell 1656), which can be replicated for any other system hosting a hot gaseous halo.
Our major conclusions are summarized as follows.
\renewcommand{\labelitemi}{$\bullet$}
\begin{itemize}
\item {
Relative 3-d pressure perturbations $\delta P/P$ can be retrieved via the 2-d SZ perturbations $\delta y/y$,
after removal of the average background profile and spherical deprojection. 
 The retrieved pressure fluctuations peak at 500 kpc, which is at
   the limit of Planck resolution, and steadily decline beyond 1
 Mpc. When combined with the decline in X-ray derived power spectrum on smaller
   scales, we can take it as an evidence for a peak in the turbulence power spectrum at
   $\approx\,$$500~\rm{ kpc}$ scales.
 The fluctuations at such dominant scale are $\delta P/P =
  33\pm 12\%$ and $74\pm19\%$ for the $15'$ and $40'$ radius region, respectively.
Because of the limited angular resolution, {\it Planck} can not resolve the power-law cascade in the spectrum, which is instead constrained by X-ray {\it Chandra} data ($A_k\propto k^{-1/3}$; Kolmogorov slope).}\\

\item {We carefully tested the impact of foreground/background contamination,
selecting external regions without evident structures, and we provided
evidence that intrinsic SZ perturbations, and not contamination, dominate our signal} within Coma cluster, in particular within $r<1\,$Mpc.\\

\item{We show how to deproject the SZ fluctuations in Fourier space. The
    impact of projection effects is to suppress the small scale power as a
    function of the SZ window function $W(\theta,z)$, which is broader than
    that of X-ray emissivity ($\propto n^2$). The 3-d to 2-d
    amplitude ratio however does not vary drastically: it is of
    order unity at Mpc scale, increasing by a factor of a few at smaller scales.} \\

\item{SZ fluctuations are a novel important tool to probe the ICM physics,
    which optimally complements the X-ray data tied to smaller tens of kpc scales.
By using the modeling and 3-d high-resolution simulations presented in Gaspari et al.~(\citeyear{Gaspari:2013,Gaspari:2014}), which relate turbulence to thermodynamic perturbations, we improve the turbulence constraints in Coma. The turbulence Mach number is ${\rm Ma_{\rm 3d} =0.8\pm0.3}$ (15' region) with injection scale $L_{\rm inj}\approx 500$ kpc. For the 40' region, the Mach number doubles, albeit velocities remain similar due to the declining plasma temperature. The transonic value at larger radii suggests that we are approaching the accretion shock region.}\\

\item{The large SZ fluctuations imply that the hot halo is in adiabatic mode (mediated by sound waves), rather than in isobaric mode (mediated by buoyancy waves). The large injection scale and velocities $\approx1.2$\,-\,$1.8\times10^3$\,km\,s$^{-1}$ are consistent with driving due to mergers, in particular tied to internal galaxy groups -- in agreement with cosmological simulations and complementary X-ray data.}\\

\item{The PDF of SZ fluctuations is log-normal with mild non-Gaussianities (heavier tails). 
This is the same statistics of perturbations predicted by high-resolution simulations of turbulence, with heavier tails induced by the increasingly compressive motions (since approaching the transonic regime).}\\

\item{The non-thermal pressure support is $E_{\rm turb}/E_{\rm
      th}\simeq0.34$ (for 15' radius), corroborating Coma belongs to the
    class of unrelaxed clusters.
    We propose a simple methodology (Eq.~\ref{e:bias}-\ref{e:bias3}) to study the mass bias $b_M=M_{\rm x}/M-1$ as a function of Mach number.
    The retrieved turbulent pressure can induce significant HSE mass bias $b_M\approx -15\%$ to $-45$\% in the core and cluster outskirts, respectively.
      The SZ fluctuations spectrum thus allows to better retrieve the true masses, improving our ability to carry out precision cosmology, in terms of measuring the cosmological parameters (e.g., $\sigma_8$) and the main scaling relations (e.g., $L_{\rm x}-M_{\rm tot}$, $M_{\rm tot}-T_{\rm x}$).}

\end{itemize}

This study opens up a new window into the diffuse gas/plasma physics.
While total SZ power conveys the total energy content, the SZ fluctuations unveil the non-thermal deviations, thus providing a self-consistent way to assess the global cluster thermodynamics.
The ground based high-resolution SZ missions -- SPT \citep{sptpol},
ALMA \citep{alma}, ACT \citep{actpol}, CARMA \citep{carma}, Mustang \citep{Mustang2} --
  will be able to extend our work based on {\it Planck} maps and assess the
  full cascade of perturbations.
  Future CMB space experiments as COrE (\citealt{core}; Cosmic Origins Explorer) 
  and PRISM (Polarized Radiation Imaging and Spectroscopy
  Mission; \citealt{prism}) may even probe the small scales accessible to X-ray, due to the better
  S/N and foreground cleaning.
  Considering that, at the present time, the next large X-ray mission {\it Athena} has foreseen launch in 2028,
  the aforementioned SZ missions are vital to advance cluster astrophysics.
  In closing, we remark the importance of combining multi-wavelength observations and multi-scale simulations,
  as probing the astrophysics of the diffuse gas and hot plasmas is inevitably tied to cosmological studies, and vice versa.

\section*{Acknowledgements}
This paper used observations obtained with {\it Planck}
(\url{http://www.esa.int/Planck}), an ESA science mission with instruments and
contributions directly funded by ESA Member States, NASA, and Canada. This
research also made  use of the HEALPix software \citep{healpix}
 (\url{http://healpix.sourceforge.net}) and \emph{Aladin sky atlas} developed at CDS,
 Strasbourg Observatory, France \citep{aladin}.
M.G. is supported by NASA through Einstein Postdoctoral Fellowship Award
Number PF-160137 issued by the Chandra X-ray Observatory Center, which is
operated by the Smithsonian Astrophysical Observatory for and on behalf of
NASA under contract NAS8-03060. 
This work made use of the computational
resources of Department of Theoretical Physics, TIFR. 
We thank D.~Eckert, N.~Battaglia,
D.~Spergel, J.~Stone, S.~Molendi, F.~Gastaldello, E.~Pointecouteau,
E.~Churazov for helpful comments. 
{We are grateful to the referee for the detailed insights which helped to improve the manuscript.} 

\bibliographystyle{mnras}

\bibliography{coma_final}

\appendix

\section{Flat sky approximation}\label{appa}
We review the flat sky approximation following \citet{jk1998}. Let us
consider the correlation function written in terms of Fourier space  as
well as in spherical harmonic space. Equating the two ways of calculating
the same correlation function gives us the flat sky relation between the
two power spectra. 

We will use angular brackets to denote ensemble average. In flat space two dimensional Cartesian position vector is
$\bx=(x \cos\phi,x\sin\phi)$, where the bold font denotes a vector and
normal font its scalar amplitude. The
correlation function for a field $f(\bx)$ for two points separated by
displacement vector $\br$ is, assuming statistical homogeneity,
\begin{align}
C(r)&\equiv \langle f(\bx+\br)f(\bx)\rangle\nonumber\\
&=\int\frac{\id^2k}{\left(2\pi\right)^2}\expe^{i\bk.\bx}
\frac{\id^2k'}{\left(2\pi\right)^2}\expe^{i\bkp.(\bx+\br)}\langle
f(\bk)f(\bkp)\rangle\nonumber\\
&=\int\frac{\id^2k}{\left(2\pi\right)^2}\expe^{-i\bk.\br}P_f(k)\nonumber\\
&=\int \frac{k\id k \id
  \phi_k}{\left(2\pi\right)^2}\expe^{-ikr(\cos\phi_k\cos\phi+\sin\phi_k\sin\phi)}P_f(k)\nonumber\\
&=\int \frac{k\id k}{2 \pi}J_0(kr)P_f(k),\label{eq:pk}
\end{align}
where we have used $\bk=(k\cos\phi_k,k\sin\phi_k)$, $\langle
f(\bk)f(\bkp)\rangle= (2\pi)^2\delta_D(\bk_\bkp)P_f(k)$, $\delta_D$ is the
Dirac delta distribution, and $P_f(k)$ is the Fourier space power
spectrum. In the last line we integrated over the Fourier angle $\phi_k$
and $J_0$ is the Bessel function of first kind.

Doing the same exercise on a sphere of radius $R$ so that $r=R\theta$, and
$\theta$ is the angular distance between the two points on the sphere, we
get
\begin{align}
C(r)=C(R\theta) =\sum_{\ell}\frac{2\ell+1}{4\pi}\mcP_{\ell}(\cos(\theta))C_{\ell}
\end{align}
Using the fact that for $\ell\gg 1$ we have
$\mcP_{\ell}(\cos(\theta))\approx J_0(\ell\theta)$ and approximating the
sum by an integral we get
\begin{align}
C(r)\approx \int \frac{\ell\id \ell}{2\pi} J_0\left(\frac{\ell r
  }{R}\right) C_{\ell}\label{eq:cl}
\end{align}
From Eqs. \ref{eq:pk} and \ref{eq:cl} we get $k^2P_f(k)\approx
\ell^2C_{\ell}|_{\ell=kR}$. Note that $k$ in our convention is the angular frequency with
$k=2\pi/\lambda$, where $\lambda$ is the physical wavelength of the Fourier
mode. We use PolSpice software package \citep{ps1,ps2} software
  package which calculates the power spectra in the spherical harmonic
  domain and use the flat sky approximation to present the results in the
  Fourier domain. Note that since we have data on a sphere, spherical
  harmonics are the natural basis for analyzing the data.

\section{Relation between pressure and SZ power spectrum assuming spherical
  symmetry}\label{appb}
We can get the relationship between the pressure and SZ power spectrum
under the assumption of spherical symmetry by calculating the correlation
function in two different ways. For the SZ correlation between two points
at $\bt$ and $\bt+\br$ we have
\begin{align}
C(r)\equiv \langle \frac{\delta y}{y}(\bt)\frac{\delta y}{y}(\bt+\br)\rangle &= 
\int \frac{\id ^2 \kt}{\left( 2\pi \right)^2}\expe^{i\bkt.\bt}\frac{\id ^2
  \ktp}{\left( 2\pi
  \right)^2}\expe^{i\bktp.(\bt+\br)}\langle\tilde{y}(\bkt)\tilde{y}(\bktp)\rangle\nonumber\\
&=\int \frac{\id ^2 \kt}{\left( 2\pi \right)^2}\expe^{-i\bkt.\br}P_y(\kt),\label{Eq:ypower}
\end{align}
where $\tilde{y}$ is the Fourier transform of $\frac{\delta y}{y}$ and we
have used 
\begin{align}
\langle\tilde{y}(\bkt)\tilde{y}(\bktp)\rangle=\left(2\pi\right)^2P_y(\kt)\delta_D^2(\bkt+\bktp),
\end{align}
and $\delta_D$ is the Dirac delta distribution.
Alternatively we can write $y$ as integral over pressure along the line of
sight direction which we take along the $z$ direction,
\begin{align}
C(r)\equiv \langle \frac{\delta y}{\bar{y}}(\bt)\frac{\delta y}{\bar{y}}(\bt+\br)\rangle &=
\langle \frac{\int \bar{P}(\delta P/\bar{P})\id z}{\int \bar{P}\id z}(\bt)\frac{\int \bar{P}(\delta P/\bar{P}) \id
  z}{\int \bar{P} \id z}(\bt+\br)\rangle
\end{align}
Defining the window function 
\begin{align}
W(z)\equiv \frac{\bar{P}}{\int \bar{P} \id z} =\frac{\me c^2}{\sigT}\frac{\bar{P}}{ \bar y } 
\end{align}
and assuming that it is independent of $\bt$, we get
\begin{align}
C(r)&=\langle \int\id z~ W \frac{\delta P}{\bar{P}}(\bt)\int \id z'W\frac{\delta P}{\bar{P}} (\bt+\br)\rangle\nonumber\\
&=\int \id z\id
z'\frac{\id^3k}{\left(2\pi\right)^3}\frac{\id^3k'}{\left(2\pi\right)^3}\frac{\id
  k_{z_1}'}{2\pi}\frac{\id
  k_{z_2}'}{2\pi}\tilde{W}\left(k_z-k_{z_1}'\right)\tilde{W}\left(k_z'-k_{z_2}'\right)\nonumber\\
&\expe^{i\left(\bkt.\bt+k_z.z+\bktp.(\bt+
\br)+k_z'.z'\right)}\langle\tilde{P}\left(\bkt,k_{z_1}'\right)\tilde{P}\left(\bktp,k_{z_2}'\right)\rangle,
\end{align}
where we have defined $\tilde{P}$ as the Fourier transform of $\delta
P/\bar{P}$ and $\tilde{W}$ as the Fourier transform of $W$. Using 
\begin{align}
\langle\tilde{P}\left(\bkt,k_{z_1}'\right)\tilde{P}\left(\bktp,k_{z_2}'\right)\rangle=\left(2\pi\right)^3\delta_D^2\left(\bkt+\bktp\right)\delta_D\left(k_{z_1}'+k_{z_2}'\right)P_p\left(\left|\bkt+\bkz_1'\right|\right)
\end{align}
and integrating using the Dirac delta distributions we get
\begin{align}
C(r)=\int \frac{\id^2 \kt}{\left(2\pi\right)^2} \frac{\id k_z}{2\pi}\left|\tilde{W}(k_z)\right|^2\expe^{i\bkt.\br}P_p\left(\left|\bkt+\bkz\right|\right)
\end{align}
Comparing  with Eq. \ref{Eq:ypower} we get the relation between the 2-d and
3-d power spectrum
\begin{align}
P_y(\kt)=\int\frac{\id k_z}{2\pi}\left|\tilde{W}(k_z)\right|^2P_p\left(\left|\bkt+\bkz\right|\right)
\end{align}
\section{Fourier convention}
We have used the Fourier convention where the reverse transform is given by
(in $n$ dimensions)
\begin{align}
f(\bx)=\int \frac{\id^n k}{(2\pi)^n} \tilde{f}(\bk)\expe^{i\bk.\bx}
\end{align}
X-ray studies typically 
use the Fourier convention (e.g., \citealt{churazov2012,zc2012}; \citealt{Gaspari:2013})
\begin{align}
f(\bx)=\int \id^n q ~\tilde{f}(\bq)\expe^{i2\pi \bq.\bx},
\end{align}
where $q=k/(2\pi)$. The power per unit logarithmic frequency interval, $A$,
is dimensionless and independent of the Fourier convention used. By
equating the real space variance, $C(0)\equiv \langle f(\bx)f(\bx)\rangle$, as in the previous
section, we get the relation between the power spectrum in the two
conventions as (with the subscript $k$ or $q$ specifying the Fourier
convention)
\begin{align}\label{eq:Ak}
A_q&=\sqrt{4\pi q^3 P_q(q)}=
A_k=\sqrt{\frac{k^3}{2\pi^2} P_k(k)}
\end{align}
in 3-d and 
\begin{align}
A_q&=\sqrt{2\pi q^2 P_q(q)}=
A_k=\sqrt{\frac{k^2}{2\pi} P_k(k)}
\end{align}
in 2-d. The Fourier conventions for the power spectrum in $n$ dimensions are
\begin{align}
\langle \tilde{f}(\bq) \tilde{f}(\bqp)\rangle =
\delta_D^n(\bq-\bqp)P_q(q)\nonumber\\
\langle \tilde{f}(\bk) \tilde{f}(\bkp)\rangle =
(2\pi)^n\delta_D^n(\bk-\bkp)P_k(k),
\end{align}
where $\delta_D^n$ is the $n$-dimensional Dirac delta distribution.
\end{document}